\documentclass[10pt,journal,compsoc]{IEEEtran}

\usepackage{amsmath,amsfonts}
\usepackage{algorithmic}
\usepackage{algorithm}
\usepackage{array}
\usepackage[caption=false,font=normalsize,labelfont=sf,textfont=sf]{subfig}
\usepackage{textcomp}
\usepackage{stfloats}
\usepackage{url}
\usepackage{verbatim}
\usepackage{graphicx}
\usepackage{cite}
\usepackage{booktabs} 
\usepackage{tabularx} 
\usepackage{threeparttable}
\usepackage{color}
\usepackage[normalem]{ulem}

\begin{document}
\title{Diff-3DCap: Shape Captioning with Diffusion Models}
\author{Zhenyu~Shu,
Jiawei~Wen*,
Shiyang~Li,
Shiqing~Xin,
Ligang~Liu% <-this % stops a space
\IEEEcompsocitemizethanks{
	\IEEEcompsocthanksitem Zhenyu Shu is with School of Computer and Data Engineering, NingboTech University, Ningbo 315100, China. He is also with Ningbo Institute, Zhejiang University, Ningbo 315100, China.
	\protect\\
	E-mail: shuzhenyu@nit.zju.edu.cn (Zhenyu Shu)
	\IEEEcompsocthanksitem Jiawei Wen is with College of Computer Science and Technology, Zhejiang University, Hangzhou, PR China. Corresponding author.
	\protect\\
	E-mail: jiaweiwen\_paper@163.com (Jiawei Wen)
	\IEEEcompsocthanksitem Shiyang Li is with College of Computer Science and Technology, Zhejiang University, Hangzhou, PR China.
	\IEEEcompsocthanksitem Shiqing Xin is with School of Computer Science and Technology, ShanDong University, Jinan, PR China.
	\IEEEcompsocthanksitem Ligang Liu is with Graphics \& Geometric Computing Laboratory, School of Mathematical Sciences, University of Science and Technology of China, Anhui, PR China.}
\thanks{Manuscript received month day, year; revised month day, year.}
}


\markboth{IEEE transactions on visualization and computer graphics,~Vol.~XX, No.~X, Month~Year}%
{Shell \MakeLowercase{\textit{et al.}}: Bare Demo of IEEEtran.cls for Computer Society Journals}



\IEEEtitleabstractindextext{%
\begin{abstract}
	% 3D shape captioning 是一个重要的任务，并且受到了越来越多的关注。然而，传统方法往往依赖于昂贵的体素表达或者目标检测算法，效果仍然不够理想。为了应对以上挑战，本文提出了一种基于
	The task of 3D shape captioning occupies a significant place within the domain of computer graphics and has garnered considerable interest in recent years. Traditional approaches to this challenge frequently depend on the utilization of costly voxel representations or object detection techniques, yet often fail to deliver satisfactory outcomes. To address the above challenges, in this paper, we introduce Diff-3DCap, which employs a sequence of projected views to represent a 3D object and a continuous diffusion model to facilitate the captioning process. More precisely, our approach utilizes the continuous diffusion model to perturb the embedded captions during the forward phase by introducing Gaussian noise and then predicts the reconstructed annotation during the reverse phase. Embedded within the diffusion framework is a commitment to leveraging a visual embedding obtained from a pre-trained visual-language model, which naturally allows the embedding to serve as a guiding signal, eliminating the need for an additional classifier. Extensive results of our experiments indicate that Diff-3DCap can achieve performance comparable to that of the current state-of-the-art methods.
\end{abstract}

}

\maketitle

\IEEEdisplaynontitleabstractindextext

% \fi
\IEEEpeerreviewmaketitle
\IEEEraisesectionheading{\section{Introduction}}
% 研究动机
\IEEEPARstart{S}{hape} understanding is crucial across multiple domains, including robotics, augmented reality, and 3D modeling. Within this field, shape captioning emerges as a challenging task. It involves the automated creation of descriptive captions for 3D shapes, thereby recognizing and understanding their geometrical and topological attributes. More importantly, shape captioning articulates these attributes in natural language, bridging the divide between geometric perception and linguistic articulation. This process augments the comprehension of 3D shapes and enhances user interaction and engagement by translating complex geometric data into accessible and understandable language.

% 已有方法
\begin{figure}[t]
	\centering
	\includegraphics[width=1\columnwidth]{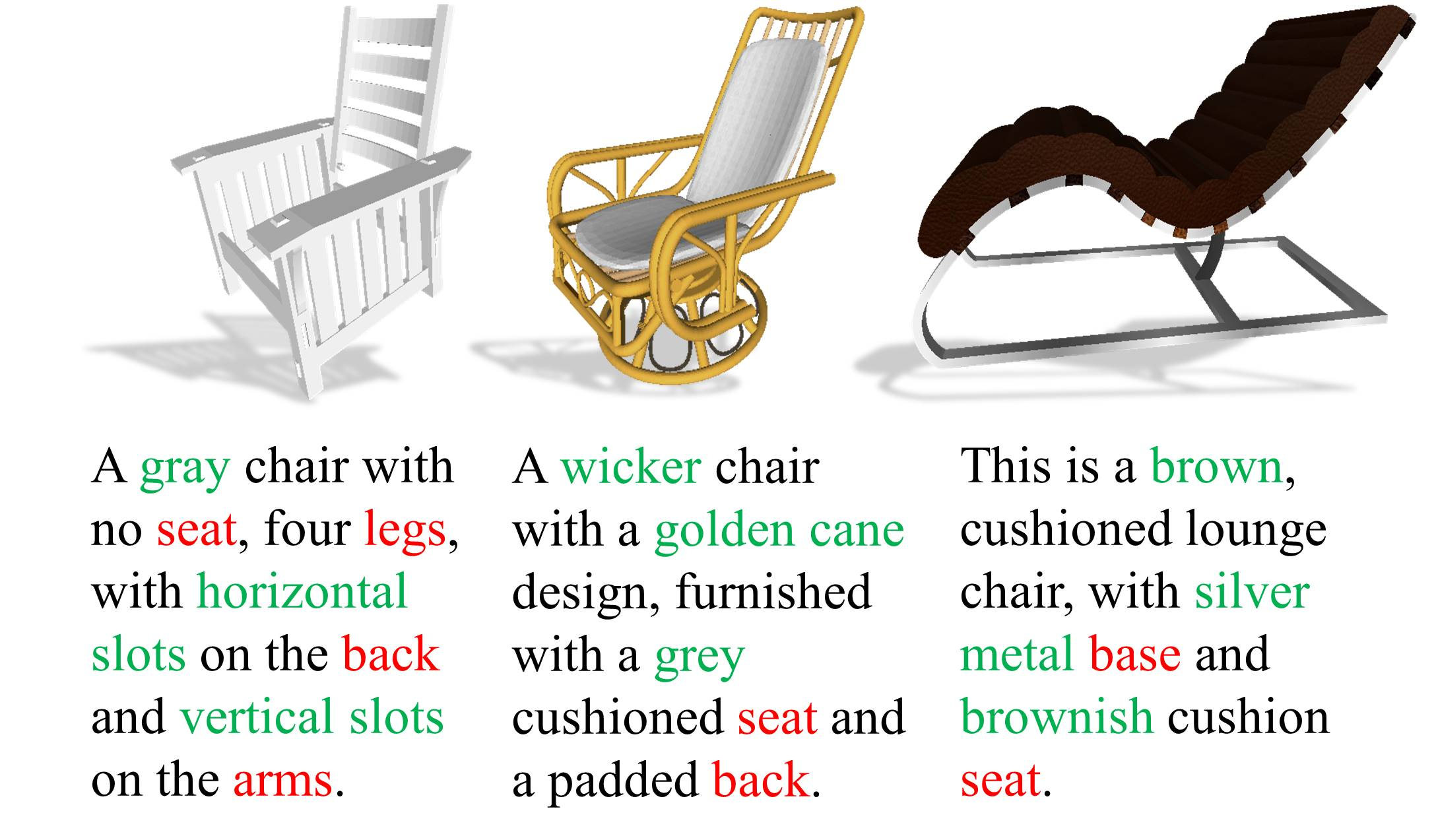}
	\caption{Captions produced by our proposed Diff-3DCap model are presented. Words colored in green signify outstanding attributes, and those in red denote the part classes of the 3D shape. This pattern aligns with human observation habits.}
	\label{fig:homepage_samples}
\end{figure}
In recent pioneering efforts, Text2Shape~\cite{chen2019text2shape} has played a critical role in 3D shape captioning, providing a valuable dataset that pairs 3D shapes with corresponding descriptive annotations. They utilized an encoder to compute global features for 3D shapes represented by voxels.
Subsequently, Han \textit{et al.}~\cite{han2019y2seq2seq} introduce a methodology that utilizes sequences of views to represent 3D shapes, effectively addressing and mitigating the challenge of high cubic complexity and facilitating greater scalability of 3D shape representation. The approach incorporates unimodal reconstruction and cross-modal prediction.
To take a further step, the methodology deployed by ShapeCaptioner~\cite{han2020shapecaptioner}, which encompasses learning the geometry of parts through segmentation benchmarks, facilitates the comprehension of semantic components within 3D shapes. The acquired insights are applied to the 3D-Text dataset, generating accurate bounding boxes of object detection for rendered views.
Moreover, Luo \textit{et al.}~\cite{luo2024scalable} broaden the research domain by focusing on a significantly more extensive 3D-text dataset named Objaverse~\cite{deitke2023objaverse}, which encloses a vast array of 3D shapes and is an order of magnitude larger than any previously available 3D-Text dataset. Besides, they employed several large-scale pre-trained models, including BLIP2~\cite{li2023blip} to generate captions for render views, CLIP~\cite{radford2021learning} to recognize captions with high quality, and GPT4~\cite{achiam2023gpt} to effectively synthesize perspective-varied information.
Recently, Liu \textit{et al.}~\cite{liu2024openshape} introduced an approach for learning joint representations of image, text, and point clouds across different modalities. To achieve superior abilities for open-world recognition, they focused on scaling up training datasets by ensembling, filtering, and enriching existing datasets. Besides, they adopted strategies for scaling up 3D backbone networks. Consequently, their refined shape representations can be integrated with CLIP-based models to facilitate point cloud captioning.

% conclusion
While these works have made significant strides in 3D shape captioning, several challenges exist. 
Firstly, due to computational resource constraints, the resolution bottleneck hampers the ability to capture fine-grained details. 
Furthermore, generating detailed part-level descriptions remains a challenge, which is essential for a comprehensive understanding of the semantic attributes of 3D shapes, aligning with human observational habits.
Moreover, the computational complexity introduced by object detection and large-scale pre-trained models may not be suitable for real-time applications or systems with limited processing power. 
Additionally, including atypical 2D projection views in larger datasets may compromise the accuracy and variety of the generated descriptions.
Consequently, capturing local features efficiently and generating results with high accuracy and wide variety remains challenging for advancing the field of 3D shape captioning.

% 本文方法概要
To address this issue, we propose Diff-3DCap, a novel approach aimed at automating the generation of captions for 3D shapes from multiple perspectives, utilizing a continuous diffusion model and a lightweight pre-trained model. The process begins by transforming 3D shapes into 2D projection views from various viewpoints. These projections and their corresponding textual captions are then processed through a pre-trained visual-language model to obtain embeddings, which serve as input for a continuous diffusion model.
The lightweight visual-language model utilizes a single-stream architecture, allowing it to avoid the complexities of handling visual and textual information separately. Additionally, the embedding model does not depend on traditional Convolutional Neural Networks or region-based methods for local feature extraction, such as object detection. Instead, it directly processes image patches through a linear embedding layer. This approach reduces computational costs and speeds up the inference process. Our model can efficiently generate visual embeddings that capture local details from rendered views by leveraging these strengths.
Furthermore, throughout the forward phase of noise infusion and the subsequent reverse phase of noise reduction, the model progressively acquires the capability to generate detailed descriptions guided by visual embeddings. This process's culmination is synthesizing these perspective-specific narratives into a comprehensive and unified caption. Figure~\ref{fig:homepage_samples} presents some generated samples of our method.

% 研究贡献，尽量从我们的工作有何创新点，解决了以往工作的什么问题.
The main contributions of our work are outlined as follows:
\begin{itemize}
    \item
    To accomplish the shape captioning task, we combine latent representations derived from rendered images and textual captions within a continuous diffusion framework, which can ensure generated results with good quality and semantic similarity.
    \item
    We take into account the local features of 3D shapes through render patches, avoiding the high training costs of voxel representation or prolonged inference time associated with object detection.
    \item
    With our efficient consolidation method, we effectively aggregate captions of different perspectives to form a comprehensive description of 3D objects.
\end{itemize}

% 文章结构
The rest of this paper is structured as follows. Section~\ref{rw} offers an exhaustive overview of the domains pertinent to our research. Section~\ref{md} outlines the entire methodology of our proposed model, with a detailed examination of its architecture. Section~\ref{exp} compares experimental outcomes with contemporary state-of-the-art techniques. Section~\ref{lf} delves into the limitations of our study and proposes directions for future research. Section~\ref{cl} provides a conclusion that encapsulates the essence of our work.

\section{Related Work}\label{rw}
This section offers an overview of the domains related to our work, including diffusion model, image captioning, and shape captioning.

\subsection{Diffusion model}
The diffusion model has emerged as a powerful tool for a variety of applications, including image generation~\cite{ho2020denoising, saharia2022photorealistic, rombach2022high, graikos2024learned, dhariwal2021diffusion}, text generation~\cite{li2022diffusion, wu2023ar, gong2022diffuseq, lin2023text, lin2022genie},  and video generation~\cite{ho2022video, ho2022imagen,kim2023diffusion, wu2023tune}.

% image generation
For image synthesis, Denoising Diffusion Probabilistic Models~\cite{ho2020denoising} introduces a novel training method leveraging a weighted variational bound, informed by the connection between diffusion probabilistic models and denoising score matching with Langevin dynamics. This method significantly improves the model's capacity to generate high-fidelity images, showcasing the potential of diffusion probabilistic models in achieving high-quality image synthesis for various applications.
AnimeDiffusion~\cite{10412677} introduces a novel approach leveraging hybrid diffusions to automatically colorize anime face line drawings. Through a two-phase end-to-end training strategy, it achieves semantic correspondence and color consistency and outperforms prevailing methods based on generative adversarial networks in qualitative and quantitative assessments.

% text generation
In the context of text generation, Li \textit{et al.}~\cite{li2022diffusion} pioneered the investigation into applying continuous diffusion models for processing discrete textual data instead of operating within a discrete state space. Furthermore, Li \textit{et al.}~\cite{li2022diffusion} examine six specific control tasks, encompassing fine-grained objectives such as semantic content modulation and more complex structural goals like adherence to syntactic parse tree constraints. Notably, several control objectives they explore do not necessitate using classifiers, including regulating sequence length and facilitating content infilling.
Gong \textit{et al.}~\cite{gong2022diffuseq} represent the pioneering effort to apply diffusion models to the SEQ2SEQ text generation task, facilitating end-to-end training without the necessity for classifiers.

% video generation
Ho \textit{et al.}~\cite{ho2022video} extend the traditional image diffusion architecture to the domain of creative video generation, allowing for training that leverages both image and video data sources simultaneously. Additionally, Ho \textit{et al.}~\cite{ho2022imagen} showcase the effectiveness and simplicity of cascaded diffusion models for generating high-resolution videos.
Moreover, Make-Your-Video~\cite{10436391} explores joint-conditional video generation through context description and temporal depth using a pre-trained Latent Diffusion Model.

\subsection{Image captioning}
% image captioning definition
Image captioning, bridging the domains of computer vision and natural language processing, aims to generate descriptive annotations for images automatically. Initial methods depended on manually designed features and conventional machine learning approaches, constraining their capacity to discern intricate image details.

The introduction of deep learning has revolutionized this field, enabling models to derive complex representations directly from data, significantly enhancing the quality and relevance of generated captions.
% encoder-decoder architecture
~\cite{anderson2018bottom, karpathy2015deep, rennie2017self, vinyals2015show} utilize an encoder-decoder framework for image captioning, where the encoder extracts visual features from the image, which are subsequently translated into natural language descriptions by a recurrent neural network~\cite{rumelhart1986learning} serving as the decoder. This architecture efficiently compresses the image content into a fixed-length vector, facilitating its transformation into a coherent natural language caption.
% object detection
Region-based Convolutional Neural Networks (R-CNN) and their advanced versions, such as Fast R-CNN~\cite{girshick2015fast} and Faster R-CNN~\cite{ren2016faster}, have played a pivotal role in the advancement of object detection and can be a critical component of the image captioning task. These models excel in identifying and localizing objects within images, laying a robust foundation for generating relevant captions. Furthermore, Lu \textit{et al.}~\cite{lu2018neural} have introduced a framework capable of producing textual descriptions that are explicitly grounded in the entities detected in images by object detectors, thereby enhancing the accuracy and relevance of image captions. 

% attention mechanism
A series of attention-based methodologies~\cite{huang2019attention, lu2017knowing, xu2015show, yao2018exploring, yao2019hierarchy} have been introduced, enhancing the ability of models to selectively concentrate on distinct regions within an image during the sequential generation of each word in a caption, which not only improves the accuracy of the captions but also makes the generation process more interpretable. Huang \textit{et al.}~\cite{huang2019attention} have developed an augmentation to the attention mechanism, integrating it within both their model's encoding and decoding stages. Similarly, Lu \textit{et al.}~\cite{lu2017knowing} introduce an adaptive attention model equipped with a visual sentinel, offering the mechanism the capacity to determine, at each generation step, whether to reference the content of the image directly or to rely on the visual sentinel.
% Transformer
The advent of the Transformer architecture has significantly propelled the field of image captioning forward. Architectures such as the Vision Transformer (ViT)~\cite{dosovitskiy2020image} and its subsequent variants have shown exceptional capability in processing images and text within a unified framework, effectively capturing spatial and semantic details. 
Li \textit{et al.}~\cite{10502235} put all filtered image segments into a visual encoder ViT to generate their embeddings, efficiently summarizing relevant visual features.
Ji \textit{et al.}~\cite{ji2021improving} introduced an approach based on a Global Enhanced Transformer encoder to further enhance the capacity for intricate multi-modal reasoning. This encoder focuses on global features that reflect the entirety of the image, guiding the decoder to generate captions of satisfactory quality.

% AR defects
Despite the effectiveness of autoregressive methods in image captioning, these models exhibit certain drawbacks. Primarily designed to generate annotations sequentially, from left to right, where the generation of a subsequent token relies on the preceding ones, they inherently suffer from prolonged inference times and an absence of parallel processing capabilities.
% NAR
In contrast, non-autoregressive methods for image captioning, as discussed by Luo \textit{et al.}~\cite{luo2023semantic} and Gao \textit{et al.}~\cite{gao2019masked}, present a paradigm that offers enhanced parallelizability, which can substantially accelerate caption generation and enrich the variety of outcomes. Specifically, Luo \textit{et al.}~\cite{luo2023semantic} introduce a novel image captioning methodology that harnesses the capabilities of diffusion models and incorporates semantic information to produce captions that are both coherent and closely aligned with the image content. Meanwhile, Gao \textit{et al.}~\cite{gao2019masked} propose a masked non-autoregressive decoding strategy aimed at addressing the challenges associated with sequential error propagation inherent in autoregressive decoding and the multimodality dilemma encountered in non-autoregressive decoding.

\subsection{Shape captioning}
The domain of 3D shape captioning represents an interaction of computer graphics and natural language processing methodologies, with the primary aim of automating the generation of descriptive narratives for 3D objects. This interdisciplinary field has attracted considerable interest for its potential applications in virtual reality, augmented reality, and assistive technologies for individuals with visual impairments. Despite the promising utility of this domain, it is burdened by the scarcity of expansive and diverse datasets requisite for practical training. Although the release of the 3D-Text dataset~\cite{chen2019text2shape} marked a significant advancement, the exploration and development within the field of 3D shape captioning remain constrained by the limited scope of available training collections.

Initially, the work of Text2Shape~\cite{chen2019text2shape} has provided a valuable resource by pairing objects from ShapeNet~\cite{chang2015shapenet} with natural language descriptions. Central to Text2Shape is its approach to creating 3D shapes from textual descriptions. This process hinges on learning joint embeddings from annotations and colored 3D objects. The method establishes implicit connections across modalities through association and metric learning techniques. This representation adeptly encapsulates the complex many-to-many relationships between the linguistic descriptions and the physical attributes of 3D shapes.
Subsequently, Han \textit{et al.}~\cite{han2019y2seq2seq} proposed an approach that utilizes sequences of views to represent 3D shapes. This methodology enables the model to process higher-resolution data while minimizing computational requirements. Their framework introduces a unique architecture comprising two interconnected $``\text{Y}"\text{-shaped}$ sequence-to-sequence models, facilitating concurrent reconstruction and prediction of sequences. This dual approach fosters a joint understanding of both visual and linguistic modalities. 
To take a further step, the methodology introduced by ShapeCaptioner~\cite{han2020shapecaptioner} enriches the 3D shape captioning process by identifying semantic components across multiple viewpoints, consolidating these elements while retaining their inherent attributes and leveraging this consolidated data to formulate detailed captions via a sequence-to-sequence model. 
Additionally, Luo \textit{et al.}~\cite{luo2024scalable} utilize pre-trained models in image captioning, image-text alignment, and large language models to aggregate captions derived from various viewpoints of a 3D shape.
Furthermore, Luo \textit{et al.}~\cite{luo2024view} focus on optimizing viewpoint selection through a diffusion-based ranking mechanism, alleviating the hallucination problem caused by atypical rendered views.
Recently, ShapeLLM~\cite{qi2024shapellm} introduced the first 3D multi-modal Large Language Model for embodied interaction, achieving universal 3D object understanding with 3D point clouds and natural language. Its key innovations include an enhanced 3D encoder leveraging multi-view image distillation to refine geometric feature extraction.

While 3D shape captioning has seen significant advancements, enabling more precise depiction of 3D objects, there still exist several challenges. Firstly, reducing the cost of computational resources while ensuring the processing of shapes at a higher resolution represents a critical objective. Moreover, efficiently computing the local features of 3D shapes to generate descriptive texts that align with human observational habits, including fine-grained details on color, texture, and material, poses a significant challenge.

Consequently, the complexity of models and the lack of diversity in generated results continue to be considerable challenges. Here, we propose Diff-3DCap, a novel methodology designed to address these hurdles by employing a continuous diffusion model and an efficient pre-trained model requiring less computational resources. This strategy is aimed at augmenting the system's ability to autonomously generate both diverse and accurate captions for 3D shapes from multiple perspectives, thereby striving to surmount the principal obstacles in the domain.

\begin{figure*}[t]
	\centering
	\includegraphics[width=2\columnwidth]{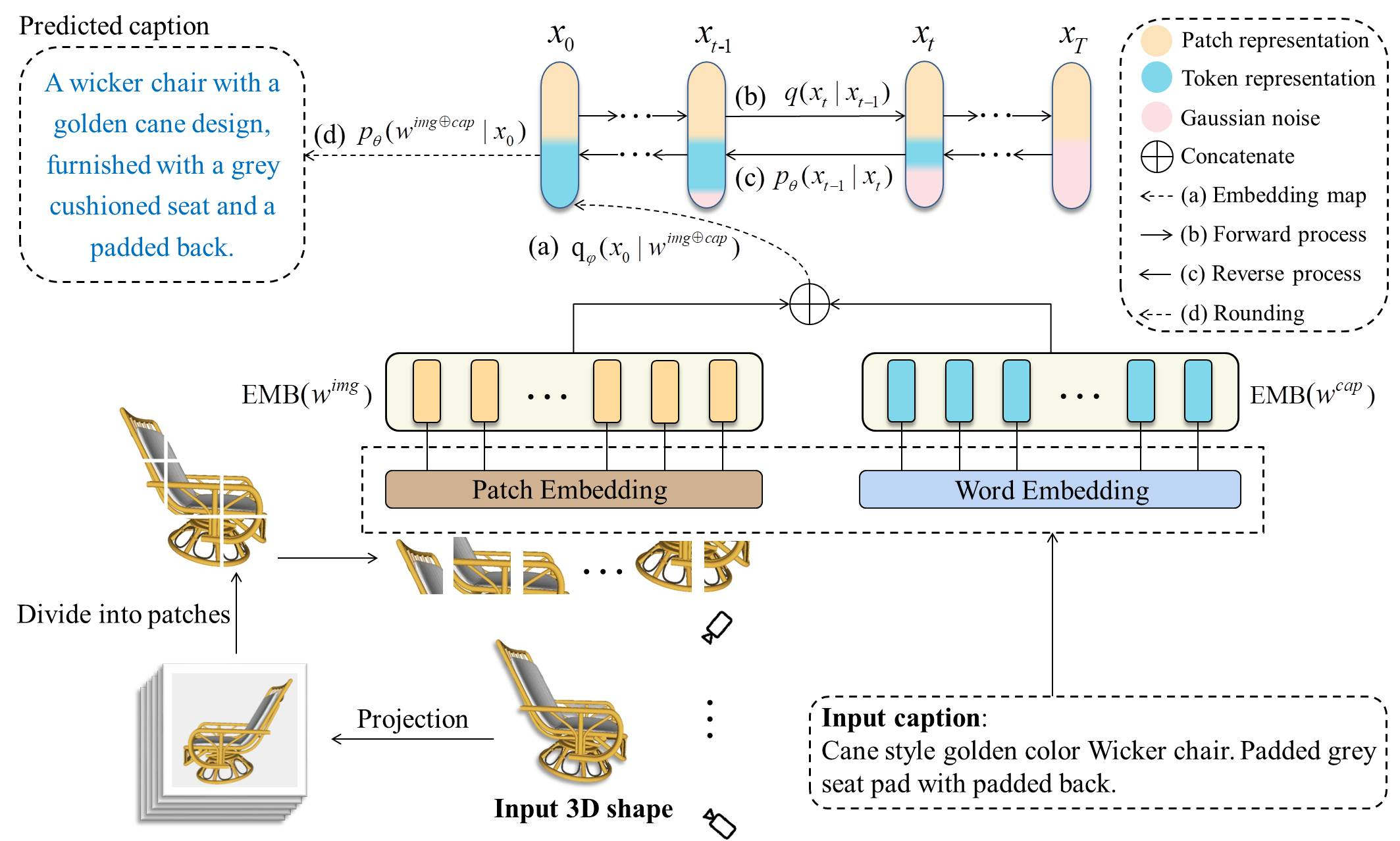}
	\caption{The demonstration of Diff-3DCap. We first render multiple 2D views from different viewpoints and then employ a pre-trained visual language model to derive text and patch embeddings. Partial Gaussian noise is added to the caption part iteratively during (b). Moreover, the reverse denoising process (c) is devoted to recovering the original state $\textit{x}_0$. Finally, the rounding process (d) executes a mapping to obtain the predicted annotation.}
	\label{fig:pipeline}
\end{figure*}
\section{Methodology}\label{md} 
We introduce Diff-3DCap, a novel approach primarily leveraging a continuous diffusion model to generate captions for 3D shapes. This methodology operates within a 3D-Text dataset~\cite{chen2019text2shape} comprising matched pairs of 3D objects and their corresponding captions.
% 具体的方法步骤
Our work adopts a four-step procedure.
First, we render multiple 2D views from different viewpoints for every 3D shape, which ensures that in each projection view, there is as much information about the entire 3D shape as possible, thus allowing the consolidated description based on all perspectives to be as comprehensive as possible.
Secondly, given the continuous nature of our diffusion model, converting discrete text data into a continuous format becomes imperative. To this end, we utilize the pre-trained visual-language model ViLT~\cite{kim2021vilt} to derive embeddings for images and captions, facilitating a seamless integration into our model's continuous space.
In the third step, our classifier-free diffusion model is introduced to propel conditional text generation, iteratively adding Gaussian noise to the caption part and denoising reversely to reconstruct the predicted caption. During the backward denoising phase, the image embedding acts as a guidance signal for text generation.
Finally, an aggregation method is employed to unify generation results from various perspectives, forming comprehensive descriptions for 3D objects.
The overview of our work is demonstrated in Figure \ref{fig:pipeline} and detailed below.

\subsection{Object rendering}
The rendering method is underpinned by a principle designed to optimize the visibility of components of 3D shapes from diverse perspectives. This strategy guarantees that the entirety of the shape's aspects are as fully captured as possible, thereby improving the accuracy of the generated descriptions and simplifying the synthesis of outcomes from various viewpoints.
Specifically, our rendering process gets a sequence of perspectives derived from an array of stationary camera points surrounding the 3D shapes at differing elevations relative to the ground plane. The initial array of viewpoints is placed precisely on the horizontal plane, the subsequent array is elevated by 30 degrees above the ground plane, and the final group is positioned 30 degrees below the horizontal plane. In our experiment, we set the number of views $V$ to 10.

\subsection{Visual language embedding}\label{sec:vl_embed}
Inspired by Diffusion-LM~\cite{li2022diffusion}, we employ an embedding function $\mathrm{EMB}(\emph{\textbf{w}})$, where $\emph{\textbf{w}}$ includes $\emph{\textbf{w}}^{cap}$ for the caption part and $\emph{\textbf{w}}^{img}$ for the image part. 
In particular, we deploy pre-trained model ViLT~\cite{kim2021vilt} instead of end-to-end dynamic trained embeddings or random Gaussian embeddings to extract the joint latent variable of image and caption concurrently, which will be set as the original status $\emph{\textbf{x}}_0$ of our continuous diffusion model.
In this way, we naturally prepare the discrete caption along with the render view for the later continuous space of our diffusion model. Concretely, when given a pair of visual and textual input data $\emph{\textbf{w}}^{img}$ and $\emph{\textbf{w}}^{cap}$, the embedding function learns their unified representation $\mathrm{EMB}(\emph{\textbf{w}}^{img\oplus cap})$ in a shared space. 
Subsequently, a transformation is applied, directing the embedding result to enter into the diffusion model framework by $q_{\phi}(\emph{\textbf{x}}_0 | \emph{\textbf{w}}^{img \oplus cap}) = \mathcal{N}(\emph{\textbf{x}}_0;\mathrm{EMB}(\emph{\textbf{w}}^{img\oplus cap}), \beta_0 \emph{\textbf{I}})$.

\subsection{Forward noising process}
At each timestep, the latent variable is denoted as $\emph{\textbf{x}}_t$, comprising $\emph{\textbf{x}}_t^{img}$ and $\emph{\textbf{x}}_t^{cap}$,  which corresponds to the image component $\emph{\textbf{w}}^{img}$ and the caption component $\emph{\textbf{w}}^{cap}$ respectively. 
Beginning with an initial latent variable $\emph{\textbf{x}}_0$ that is derived from the ground truth distribution, the forward process progressively injects noise into the caption segment $\emph{\textbf{x}}_t^{cap}$. This process continues until, at the final timestep $T$, the caption segment's latent variable conforms to a standard Gaussian distribution, denoted as $\emph{\textbf{x}}_T^{cap} \sim \mathcal{N}(0, \emph{\textbf{I}})$.
The strategy of partial noising, as inspired by~\cite{gong2022diffuseq}, involves the selective perturbation of only half of the representation. This methodology permits the remaining half to naturally serve as a guiding signal for generating captions during the denoising phase. In contrast to the work of Gong \textit{et al.}~\cite{gong2022diffuseq}, our method diverges by incorporating a multi-modal processing framework instead of focusing exclusively on unimodal SEQ2SEQ tasks.
Notably, the forward process is characterized by the absence of trainable parameters, and the transition from $\emph{\textbf{x}}_{t-1}$ to $\emph{\textbf{x}}_{t}$ is dictated by a predefined rule:

\begin{equation}
    q(\emph{\textbf{x}}_t | \emph{\textbf{x}}_{t-1}) = \mathcal{N}(\emph{\textbf{x}}_t;\sqrt{1-\beta_t}\emph{\textbf{x}}_{t-1}, \beta_t \emph{\textbf{I}}),
\end{equation}
with timestep $t \in \{1, 2,...T\}$ and the corruption hyperparameter $\{{\beta_t \in (0,1)}\}_{t=1}^{T}$, which controls the perturbation intensity of each forward step, we use a square-root noise schedule following~\cite{li2022diffusion}. Moreover, set $\alpha_t = 1 - \beta_t$ and $\overline{\alpha}_t = \prod \limits_{i=1}^{t}{\alpha_i}$, in this way, we can denote the forward noising transformation as:

\begin{subequations}\label{eq:2}
	\begin{align}
		\emph{\textbf{x}}_t &= \sqrt{1-\beta_t}\emph{\textbf{x}}_{t-1} + \sqrt{\beta_t}\epsilon_{t-1}, \\
		&= \sqrt{1-\beta_t}(\sqrt{1-\beta_{t-1}}\emph{\textbf{x}}_{t-2} + \sqrt{\beta_{t-1}}\epsilon_{t-2})+\sqrt{\beta_t}\epsilon_{t-1}, \\
            &= ... =  \sqrt{\overline{\alpha}_t}\emph{\textbf{x}}_0 + \sqrt{1-\overline{\alpha}_t}\epsilon,
	\end{align}
\end{subequations}
with Gaussian noise $\epsilon \sim \mathcal{N}(0,1)$, so intuitively $\emph{\textbf{x}}_t$ can be computed relying on $\emph{\textbf{x}}_0$ and solely once noise sampling from Gaussian distribution, 

\begin{equation}
    q(\emph{\textbf{x}}_t | \emph{\textbf{x}}_0) = \mathcal{N}(\emph{\textbf{x}}_t; \sqrt{\overline{\alpha}_t}\emph{\textbf{x}}_0, (1-\overline{\alpha}_t)\emph{\textbf{I}}).
\end{equation}

Moreover, we compute the predicted mean of Gaussian posterior $q(\emph{\textbf{x}}_{t-1}|\emph{\textbf{x}}_t,\emph{\textbf{x}}_0 )$ to simplify our later optimization objective. According to the Bayes' rule:

\begin{equation}
    q(\emph{\textbf{x}}_{t-1}|\emph{\textbf{x}}_t,\emph{\textbf{x}}_0 ) = q(\emph{\textbf{x}}_{t}|\emph{\textbf{x}}_{t-1},\emph{\textbf{x}}_0 ) \frac{q(\emph{\textbf{x}}_{t-1} | \emph{\textbf{x}}_0)}{q(\emph{\textbf{x}}_{t} | \emph{\textbf{x}}_0)},
\end{equation}
where we replace with Equation~\ref{eq:2} and obtain the mean of  $q(\emph{\textbf{x}}_{t-1}|\emph{\textbf{x}}_t,\emph{\textbf{x}}_0 )$ articulated as below,

\begin{equation}
\label{mean}
    \mu_t(\emph{\textbf{x}}_t,\emph{\textbf{x}}_0) = \frac{\sqrt{\alpha_t} (1-\overline{\alpha}_{t-1})}{1-\overline{\alpha}_{t}}\emph{\textbf{x}}_t + \frac{\sqrt{\overline{\alpha}_{t-1}}\beta_t}{1-\overline{\alpha}_t}\emph{\textbf{x}}_0. 
\end{equation}

In conclusion, we extend the partial-noising strategy on the diffusion model to the multi-modal shape captioning field, distinctly from prior uni-modal approaches. By selectively injecting noise only into the textual latent variables while preserving the image component, we enable the image features to naturally guide and constrain caption generation during the following denoising phase, which can effectively enhance generated captions' semantic coherence.
\subsection{Reverse denoising process}
Initiating with the random Gaussian noise $\emph{\textbf{x}}_T^{cap}$ about the caption segment, concatenated with the ground truth $\emph{\textbf{x}}_T^{img}$ of the image segment, the reverse denoising process endeavors to reconstruct the original textual fragment $\emph{\textbf{x}}_0^{cap}$. This process can be illustrated as follows:
\begin{subequations}
    \begin{align}
        &p_{\theta}(\emph{\textbf{x}}_{0:T}) := p(\emph{\textbf{x}}_T) \prod\limits_{t=1}^T p_{\theta}(\emph{\textbf{x}}_{t-1}|\emph{\textbf{x}}_{t}),\\
        &p_{\theta}(\emph{\textbf{x}}_{t-1}|\emph{\textbf{x}}_{t}) = \mathcal{N}(\emph{\textbf{x}}_{t-1};\mu_{\theta}(\emph{\textbf{x}}_{t}, t), \sigma_{\theta}(\emph{\textbf{x}}_{t}, t)).
    \end{align}
\end{subequations}

Specifically, our denoising architecture is based on a transformer architecture, denoted as $f_{\theta}(\emph{\textbf{x}}_t, t)$, upon which the previously mentioned $p_{\theta}$ is predicated.
Furthermore, $\mu_{\theta}(\cdot)$ and $\sigma_{\theta}(\cdot)$ represent the predicted distribution parameters of $q(\mathbf{x}_t | \mathbf{x}_{t-1})$ during the forward noising process. Particularly, the formulation of $\mu_{\theta}(\cdot)$ is akin to that presented in Equation~\ref{mean}, which is conditioned upon the observed data $\mathbf{x}_t$ as well as the predicted state of $\mathbf{x}_0$.

% objective function
To guarantee the quality of the generated captions, we employ an alternative variational lower bound to minimize the negative log-likelihood. This is formally represented as $\mathbb{E}[-\log{p_{\theta}(\mathbf{x}_0)}] \leq \mathcal{L}_{\mathrm{VLB}}$,

\begin{subequations}
\label{vlb}
    \begin{align}
        \mathcal{L}_{\mathrm{VLB}} 
    &= \mathcal{L}_T + \mathcal{L}_{T-1} + ... + \mathcal{L}_{1} + \mathcal{L}_{0}, \\
    &= \mathbb{E}_{q(\emph{\textbf{x}}_{1:T} | \emph{\textbf{x}}_0)}[\underbrace{\vphantom{log\frac{q(\emph{\textbf{x}}_{T}|\emph{\textbf{x}}_0 )}{p_{\theta}(\emph{\textbf{x}}_T)}}log\frac{q(\emph{\textbf{x}}_{T}|\emph{\textbf{x}}_0 )}{p_{\theta}(\emph{\textbf{x}}_T)}}_{\text{prior matching term}} - \underbrace{\vphantom{log\frac{q(\emph{\textbf{x}}_{T}|\emph{\textbf{x}}_0 )}{p_{\theta}(\emph{\textbf{x}}_T)}}logp_{\theta}(w^{img \oplus cap} | \emph{\textbf{x}}_0)}_{\text{reconstruction term}} \\ 
    & + \underbrace{\sum_{t=2}^{T}{log\frac{q(\emph{\textbf{x}}_{t-1} | \emph{\textbf{x}}_{t}, \emph{\textbf{x}}_{0} )}{p_{\theta}(\emph{\textbf{x}}_{t-1} | \emph{\textbf{x}}_{t})}} + log\frac{q_{\phi}(\emph{\textbf{x}}_0 | \emph{\textbf{w}}^{img \oplus cap})}{p_{\theta}(\emph{\textbf{x}}_{0} | \emph{\textbf{x}}_{1})}}_{\text{denoising matching term}}], 
    \end{align} \\
\end{subequations}
where the first prior matching term is based on the prior assumption that at the final time step, latent variable $\emph{\textbf{x}}_T$ adheres to a standard Gaussian distribution. Subsequently, the second reconstruction term measures the effectiveness of generating desirable text according to the predicted initial state $\emph{\textbf{x}}_0$, which is also the part that the rounding process mainly aims to optimize. Moreover, the remaining term utilizes the $q(\emph{\textbf{x}}_{t-1} | \emph{\textbf{x}}_{t}, \emph{\textbf{x}}_{0} )$ during forward noise injection as a supervisory signal for the backward noise removal process $p_{\theta}(\emph{\textbf{x}}_{t-1} | \emph{\textbf{x}}_{t})$, which is instrumental in enhancing the model's ability to capture the intrinsic data distribution more accurately.

In detail, for KL divergence referring to the denoising term previously discussed in Equation~\ref{vlb}, we calculate the discrepancy between two probability distributions by evaluating the divergence of their means, as specified in Equation~\ref{mean}. Furthermore, this is equivalent to assessing the gap between the predicted value of $f_{\theta}(\emph{\textbf{x}}_t, t)$ and the ground truth $\emph{\textbf{x}}_0$. This methodology follows the practice of~\cite{li2022diffusion}, which directs the model to learn the structure of $\emph{\textbf{x}}_0$ so that it can accurately align with a particular embedding.
Consequently, the $\mathcal{L}_{\mathrm{VLB}}$ can be further simplified as follows:
\begin{subequations}
\label{simple_VLB}
\begin{align}
    &\min_{\theta} 
    \mathcal{L}_{\mathrm{VLB}} 
    = \min_{\theta}\bigg[||\mathrm{EMB}(w^{img \oplus cap}) - f_{\theta}(\emph{\textbf{x}}_1, 1)||^2\\
    & + \sum_{t=2}^{T}{||\emph{\textbf{x}}_0 - f_{\theta}(\emph{\textbf{x}}_t, t)||^2}
    - logp_{\theta}(w^{img \oplus cap} | \emph{\textbf{x}}_0)\bigg], \\
    &\rightarrow \min_{\theta} 
    \bigg[||\mathrm{EMB}(w^{cap}) - \hat{f}_{\theta}(\emph{\textbf{x}}_1, 1)||^2\\
    & + \sum_{t=2}^{T}{||\emph{\textbf{x}}_0^{cap} - \hat{f}_{\theta}(\emph{\textbf{x}}_t, t)||^2} + 
    \mathcal{R}(||\emph{\textbf{x}}_0||^2)\bigg],
\end{align} \\
\end{subequations}
where the term $\mathcal{R}(\cdot)$ refers to the $L_2$ regularization constraint, and we use the $\hat{f}_{\theta}(\cdot)$ to denote the predicted text segment.
Although our objective function exclusively concentrates on textual data, the denoising transformer network structure operates on a latent variable encompassing textual and visual information. Therefore, the image influences the prediction of the text segment, and the parameters update during the backpropagation process can also affect the handling of the image.

\subsection{Caption aggregation}
\begin{figure}[t]
	\centering
	\includegraphics[width=1\columnwidth]{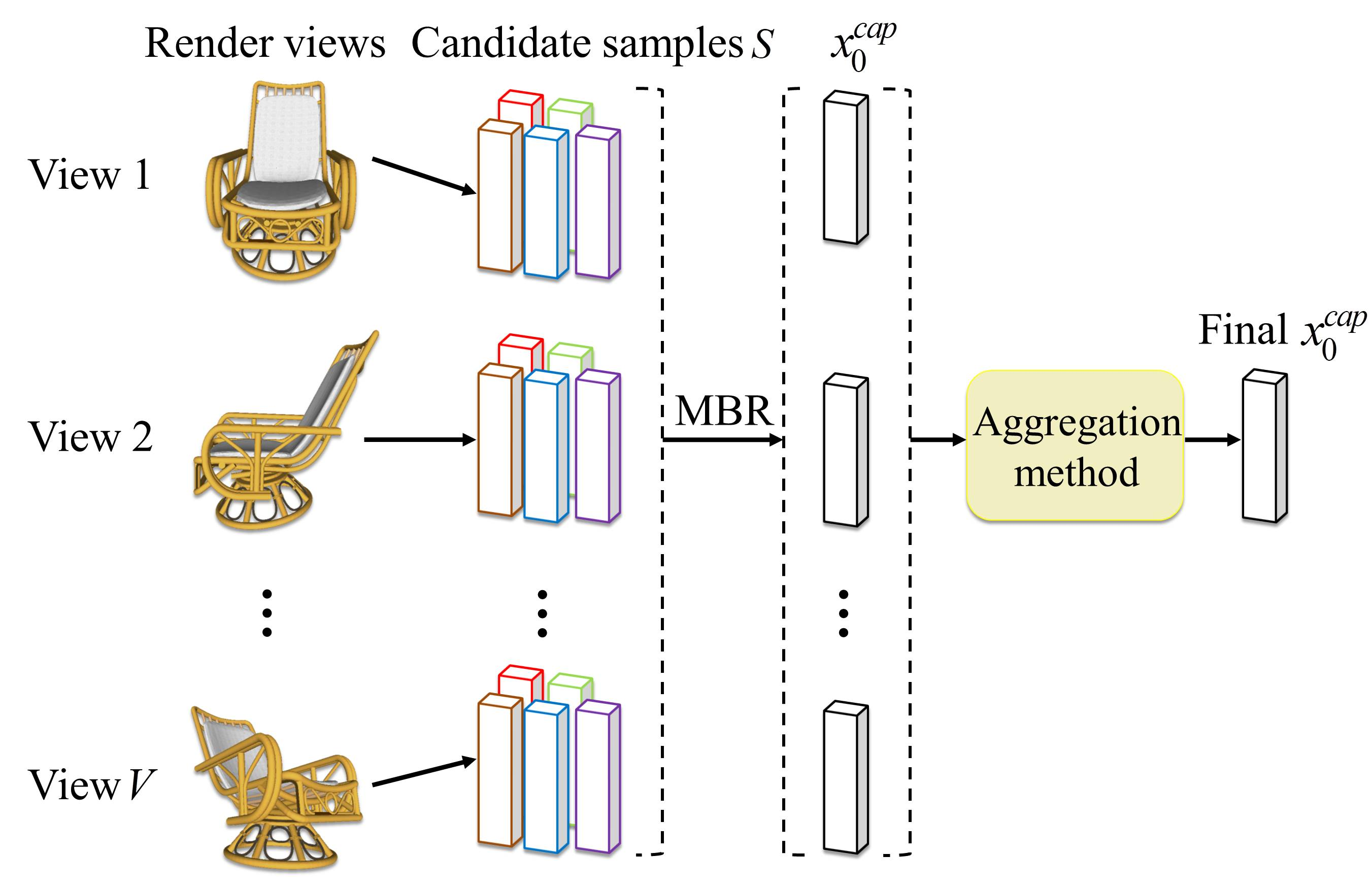}
	\caption{We generate $\textit{S}$ candidate samples for each 2D projection view, utilize Minimum Bayes Risk(MBR) to select the caption with the highest quality, and employ an aggregation method to consolidate across all perspectives for obtaining the final representation $\emph{\textit{x}}_0^{cap}$.}
	\label{fig:agg}
\end{figure}
In our methodology, we round the predicted $\emph{\textit{x}}_0$ back to a discrete caption after the inverse denoising process. Specifically, position-wise maximum likelihood estimation achieves the rounding process, which means we choose the most probable and contextually appropriate word at each position of the generated caption.
Furthermore, synthesizing generational information from diverse perspectives is challenging and essential for attaining a holistic representation of 3D objects.
% 对一个视图生成的五条候选文本进行筛选
As depicted in Figure~\ref{fig:agg}, to refine the quality of the generated captions, we adopt the broadly employed Minimum Bayes Risk (MBR) decoding strategy~\cite{kumar2004minimum}. For each projection view of the 3D shape, our approach begins with generating a set of candidate captions, denoted as $S$. Subsequently, we select a caption from $S$ that exhibits the highest quality to minimize the expected risk under a predefined loss function $\mathcal{L}$.

\begin{equation}
    \hat{\emph{\textbf{w}}_i} = argmin{_{\emph{\textbf{w}}_i \in S_i}\sum_{\emph{\textbf{w}}_i^{\prime} \in S_i} \frac{1}{|S_i|}\mathcal{L}(\emph{\textbf{w}}_i, \emph{\textbf{w}}_i^{\prime}) },
\end{equation}
where $i$ denotes the $i$-th render view of a 3D object and $\mathcal{L}$ means negative BLEU score metrics in our implementation.

% 对不同视图的最佳文本进行综合
Subsequently, the Diff-3DCap synthesizes the latent state of $\emph{\textbf{x}}_{0, i}^{cap}$ corresponding to the most outstanding $\emph{\textit{x}}_0^{cap}$ across all projection views, with $i \in \{1, 2,...,\textit{V}\}$. In practice, we aggregate all latent embeddings using a max pooling operation into a unified representation, which is then processed by the final rounding mechanism to produce a caption.

\section{Experiments}\label{exp}
In this section, we evaluate the performance of our Diff-3DCap model. First, we compare our experimental results with those of state-of-the-art approaches. Second, we investigate the impact of various hyperparameter settings on the performance of our model. Finally, we conduct ablation studies to ascertain the validity of particular network components.

\begin{figure*}[t]
	\centering
	\includegraphics[width=2\columnwidth]{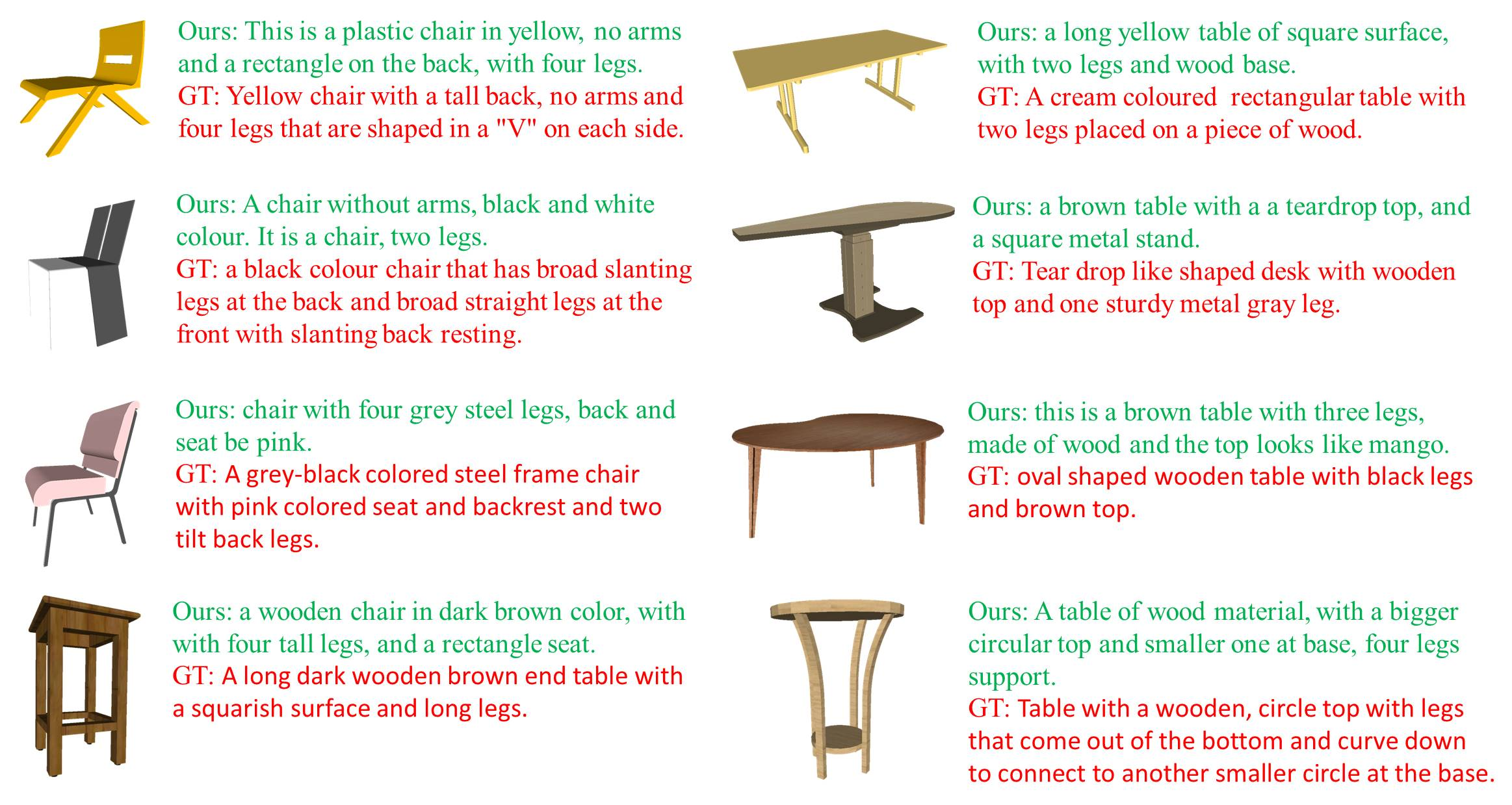}
	\caption{Comparison between results generated by our Diff-3DCap and the ground truth captions on the ShapeNet dataset. It can be observed that our model is capable of generating descriptive annotations that capture both the part components and the relevant attributes.}
	\label{fig:shapenet_res}
\end{figure*}
\begin{table*}
	\centering
	\renewcommand\arraystretch{1.2}
	\begin{threeparttable}
		\caption{The comparison of completeness metrics across different algorithms on the cross-modal 3D-Text dataset~\cite{chen2019text2shape}. A higher score denotes better performance.}
		\label{tb:shapenet_res}
		\begin{tabularx}{\textwidth}{*{10}{>{\centering\arraybackslash}X}}
			\toprule
			Method  &METEOR $\uparrow$ 			  &ROUGE $\uparrow$  				&CIDEr $\uparrow$   			  &BLEU@1 $\uparrow$   	   &BLEU@2 $\uparrow$   	&BLEU@3 $\uparrow$   	 &BLEU@4 $\uparrow$     		     \\
			\midrule
			SLR    & 0.110          & 0.240 & 0.050 		  & 0.400 & 0.170 & 0.080 & 0.040  \\
			GIF2T         & 0.160 & 0.360          & 0.140 		  & 0.610 & 0.350 & 0.210 & 0.120   \\
			V2T   & 0.210 & 0.450          & 0.270 	      & 0.670 & 0.430 & 0.260 & 0.150   \\
			SandT     & 0.209          &0.381  & 0.301		  & 0.494 & 0.338 & 0.251 & 0.214  \\	
			$\text{Y}^2\text{Seq2Seq}$       & 0.300 		  & 0.560          & 0.720 		  & 0.800 & 0.650 & 0.540 & 0.460   \\
			ShapeCaptioner   & 0.456      & 0.756  & 1.444    & 0.899 		   & 0.836         &0.785 & 0.749   \\
			OpenShape	& 0.499      & 0.758  & 1.502         &\textbf{0.904}  & 0.839         & \textbf{0.790} & 0.752   \\
			Ours         &\textbf{0.502}  & \textbf{0.760}	& \textbf{1.507} &0.825   &\textbf{0.843} & 0.779 & \textbf{0.758}   \\
			\bottomrule
		\end{tabularx}
	\end{threeparttable}
\end{table*}

\subsection{Implementation detail}
\textbf{Dataset and metrics.}
% dataset-ShapeNet
Firstly, we evaluate Diff-3DCap under a cross-modal 3D-Text dataset proposed by~\cite{chen2019text2shape}, which contains pairs of 3D shapes and corresponding captions from the ShapeNet subset~\cite{chang2015shapenet} and artificial primitives dataset. We solely use the ShapeNet subset, which includes 15,038 shapes and 75,344 captions for Tables and Chairs categories.
We adopt the same process for splitting our dataset into training and testing datasets as described in~\cite{chen2019text2shape}. Particularly, for the Tables category, we utilize 7,592 samples for training and 851 samples for testing. Meanwhile, the Chairs category includes 5,954 samples for training and 641 samples for testing.
% dataset-Cap3D
Moreover, given that the diversity of the dataset is crucial for training a captioning model with robust generalizability, we also evaluate our model on the Cap3D benchmark~\cite{luo2024scalable}, a 3D-Text dataset specifically designed for Objaverse~\cite{deitke2023objaverse}, which has a broader range of categories than ShapeNet.

% metrics for ShapeNet
For ShapeNet subset, we evaluate the Diff-3DCap model based on two aspects: completeness and semantic similarity. 
The assessment of caption completeness is conducted using metrics traditionally applied to 3D shape captioning tasks. Additionally, we compute similarity scores to ascertain the semantic coherence between the model-generated captions and the reference sentences.
For a comprehensive evaluation, we employ a suite of metrics,
including METEOR~\cite{banerjee2005meteor}, ROUGE~\cite{lin2004rouge}, CIDEr~\cite{vedantam2015cider}, 
BLEU~\cite{papineni2002bleu}, 
CLIPScore and RefCLIPScore~\cite{hessel-etal-2021-clipscore}, and
BERTScore~\cite{Zhang*2020BERTScore:}.
Specifically, the metrics utilized in our evaluation are denoted as follows:
$``\text{METEOR}"$, $``\text{ROUGE}"$, $``\text{CIDEr}"$,$``\text{BLEU@1}"$, $``\text{BLEU@2}"$, $``\text{BLEU@3}"$, $``\text{BLEU@4}"$, $``\text{CLIP-S}"$,
$``\text{Ref-CLIP}"$, $``\text{P-Bert}"$.

% metrics for Cap3D
For Cap3D dataset, in order to complement traditional linguistic overlap metrics such as BLEU, ROUGE, and METEOR, which focus primarily on surface-level n-gram matches, we incorporate several metrics including Sentence-BERT, SimCSE, and ROUGE-L. These techniques further facilitate the evaluation of semantic fidelity by assessing alignment within the embedding space.

\begin{figure*}[t]
	\centering
	\includegraphics[width=2\columnwidth]{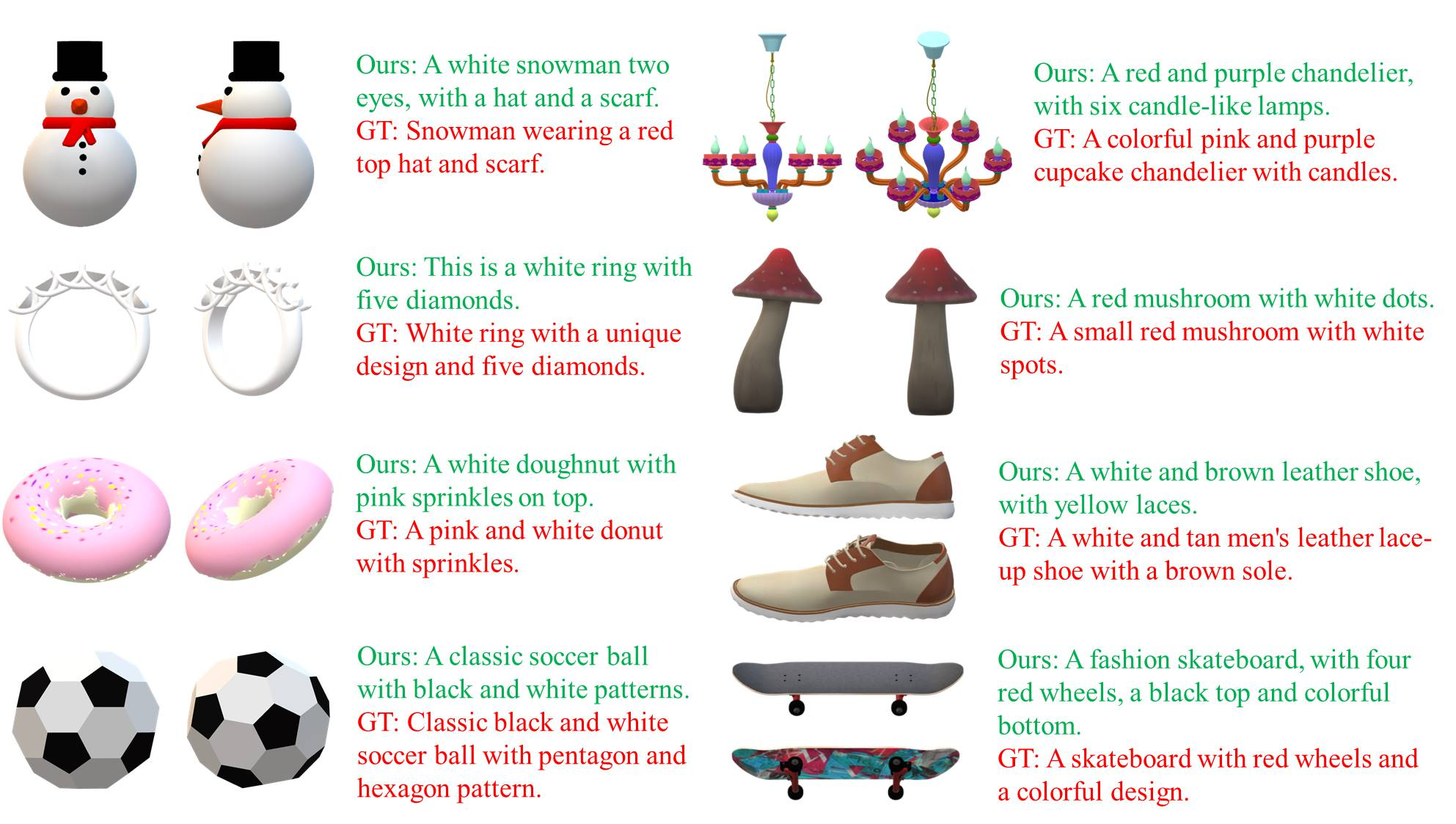}
	\caption{Comparison between results generated by our Diff-3DCap and the ground truth captions on the Cap3D dataset.}
	\label{fig:cap3d_res}
\end{figure*}

\begin{table}
	\centering
	\begin{threeparttable}
		\caption{Quantitative comparison on the Cap3D dataset. A higher score denotes better performance.}
		\label{tb:cap3d_res}
		\begin{tabularx}{\columnwidth}{
				% *4{>{\centering\arraybackslash}X}
				>{\centering\arraybackslash\hsize=1.12\hsize}X % 加宽Method列
				>{\centering\arraybackslash\hsize=1.18\hsize}X
				>{\centering\arraybackslash\hsize=0.8\hsize}X
				>{\centering\arraybackslash\hsize=0.9\hsize}X
			}
			\toprule
			Method  &Sentence-BERT $\uparrow$  &SimCSE $\uparrow$	&ROUGE-L $\uparrow$ \\
			\midrule
			3D-LLM   & 41.47  & 40.84 & 19.40 		  \\
			ShapeLLM & 45.23 & 46.91 & 19.92   \\
			GPT-4o   &\textbf{51.50} & \textbf{53.80} & 20.59   \\
			OpenShape  & 38.71 & 37.82 & 20.33  \\	
			Ours       & 40.10 & 39.23 & \textbf{21.99}  \\
			\bottomrule
		\end{tabularx}
	\end{threeparttable}
\end{table}

\textbf{Experimental setup.}
Our model, Diff-3DCap, utilizes 2D projection views, where the number of views ($V$) is established at 10. We set the dimensionality of both word and image embeddings ($H$) to 128, ensuring sufficient capacity to capture the details of captions and render views effectively. 
The model operates through a diffusion process of $T = 2000$ steps, adopting a square-root schedule to inject noise. A transformer architecture underpins the reverse denoise processing. Notably, we simplify the denoising process to enhance efficiency during inference time by reducing the steps to $T = 200$. This adjustment significantly shortens inference costs while still yielding generation outcomes that are robust and of high quality.
Moreover, we generate a batch of candidate samples $|S|=5$ for each viewpoint and then adopt the MBR strategy to select the caption with the best quality.
Ultimately, we consolidate the best captions from different perspectives by pooling the predicted original embeddings and decoding them into a final sentence.

The configurations have been meticulously selected to guarantee that our model maintains effectiveness and productivity throughout the training and inference processes.

\begin{figure}[t]
	\centering
	\includegraphics[width=0.8\columnwidth]{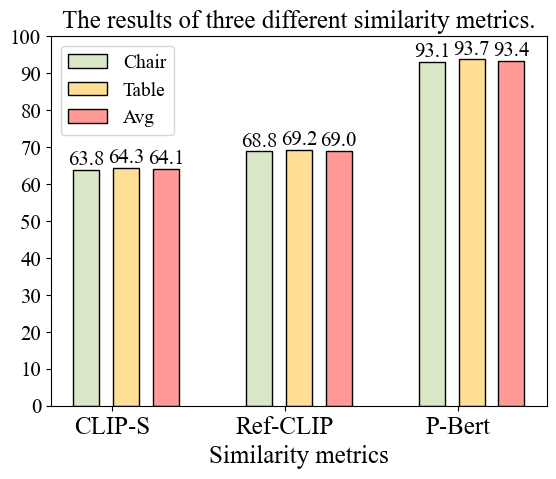}
	\caption{The performance of chair and table classes separately and the average outcome upon integrating both categories.}
	\label{fig:sim_metrics}
\end{figure}

\subsection{Comparison}
% 简单讲述下方法,衔接到实验结果
As illustrated in Section~\ref{md}, integrating the learned embeddings of captions and projected images into the continuous diffusion model facilitates a process characterized by the injection and removal of Gaussian noise. This process ultimately yields the forecasted annotation.
% 展示生成文本效果
In Figure~\ref{fig:shapenet_res}, we showcase a selection of objects about two distinct categories within the ShapeNet subset: Tables and Chairs, including both their associated predicted captions and the ground truth captions. 
Meanwhile, as depicted in Figure~\ref{fig:cap3d_res}, we present captioning results across diverse categories in the Cap3D benchmark, such as snowman, ring, doughnut, soccer ball, among others.
Our approach effectively conveys the semantic content of the objects through an explicit narrative style.

% 和其他SOTA做横向比较
To validate the effectiveness of our approach and the quality of the generated output, we performed multiple comparisons with the state-of-the-art methods in relevant fields, including SLR~\cite{shen2018sequence} for video understanding,
GIF2T~\cite{song2018cross} for cross-modal retrieval,
V2T~\cite{venugopalan2015sequence} for video captioning, SandT~\cite{vinyals2015show} for image captioning, 
$\text{Y}^2\text{Seq2Seq}$~\cite{han2019y2seq2seq} for 3D shape understanding, ShapeCaptioner~\cite{han2020shapecaptioner} for 3D shape captioning, OpenShape~\cite{liu2024openshape} for learning multi-modal joint representations,
3D-LLM~\cite{hong20233d} and ShapeLLM~\cite{qi2024shapellm} for utilizing large language models to 3D-related tasks,
and the commercial general system GPT-4o~\cite{hurst2024gpt}.

\begin{table*}
	\centering
	\renewcommand\arraystretch{1.2}
	\begin{threeparttable}
		\caption{The comparison of shape captioning results of our model using different numbers of views $V$. Here, $H=128, |S|=5$.}
		\label{tb:V}		
		\begin{tabularx}{\textwidth}{*{10}{>{\centering\arraybackslash}X}}
			\toprule
			$V$ &METEOR $\uparrow$ 			  &ROUGE $\uparrow$  				&CIDEr $\uparrow$   			  &BLEU@1 $\uparrow$   	   &BLEU@2 $\uparrow$   	&BLEU@3 $\uparrow$   	 &BLEU@4 $\uparrow$     		     \\
			\midrule
			2    & 0.132          & 0.306           & 0.018 		  & 0.329       & 0.217     & 0.185      & 0.078  \\
			4    & 0.240         & 0.489            & 0.637 		  & 0.691        & 0.369    & 0.414       & 0.242   \\
			6   & 0.335          & 0.497            & 0.792 	      & 0.721         & 0.457    & 0.518      & 0.477   \\
			8     & 0.356          &0.637          & 0.826		      & 0.776        & 0.706    & 0.658      & 0.628  \\
			10   &\textbf{0.502}  & \textbf{0.760} 	& \textbf{1.507} & \textbf{0.825}&\textbf{0.843} &\textbf{0.779} &\textbf{0.758}    \\
			\bottomrule
		\end{tabularx}
	\end{threeparttable}
\end{table*}
\begin{figure}[t]
	\centering
	\includegraphics[width=1\columnwidth]{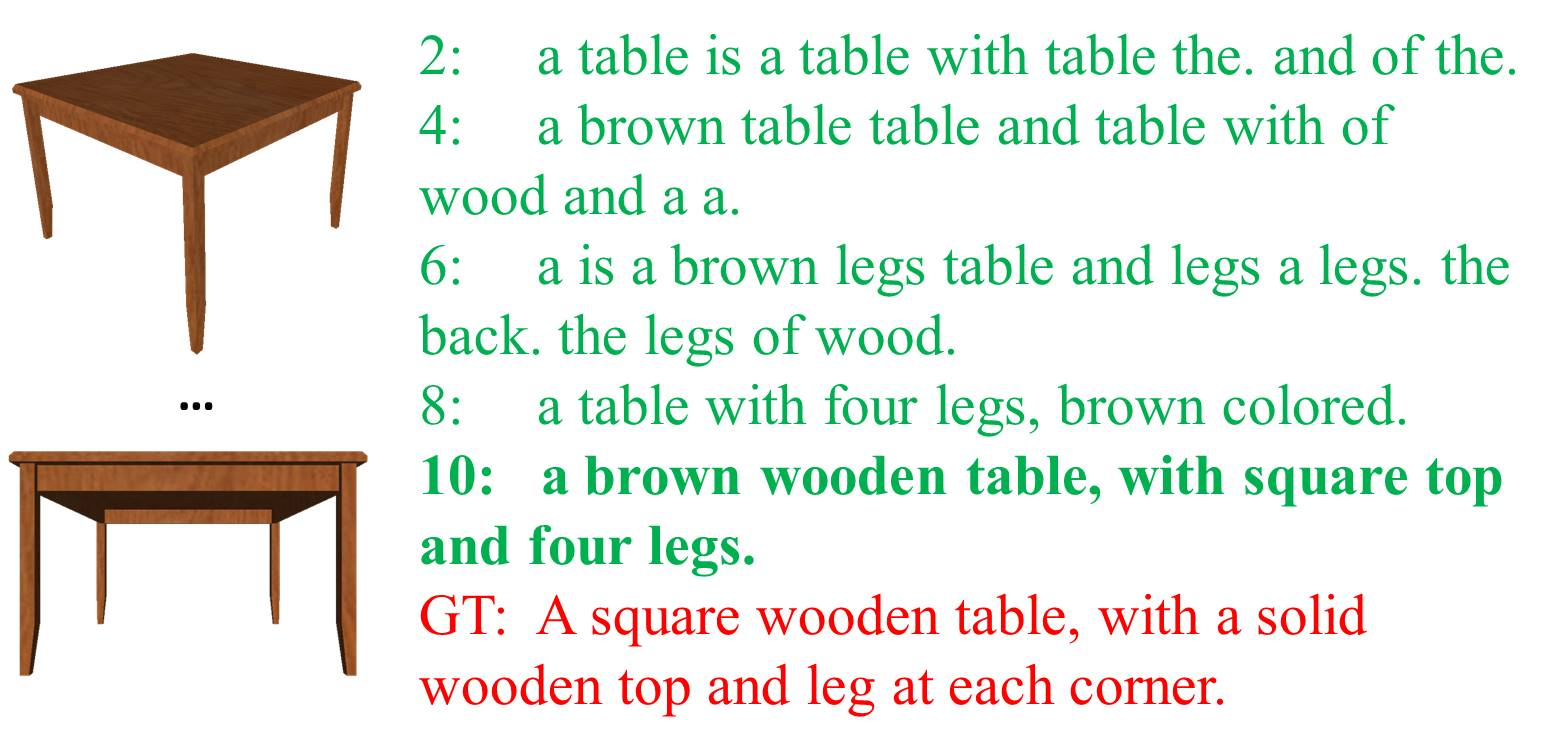}
	\caption{Comparison between generated annotations under different numbers of projection views.}
	\label{fig:ablation_view}
\end{figure}

% 所使用指标，从完备性和相似度两个角度评估
We analyzed various evaluation metrics over the ShapeNet dataset, including common metrics for text completeness and additional measures to assess the semantic similarity of predicted captions and the ground truth captions. The outcomes validate the potential of the diffusion model to 3D shape captioning tasks.
% 分析指标高或低
As presented in Table~\ref{tb:shapenet_res}, our Diff-3DCap model demonstrates good performance across nearly all evaluated metrics, with the exceptions being the $\text{BLEU@1}$ and $\text{BLEU@3}$ scores. This observation may be attributed to the inherent nature of the diffusion model, which tends to generate outputs with greater diversity. This characteristic may adversely affect the $\text{BLEU}$ scores, which are calculated based on the sequences overlapping between the predicted captions and the ground truth captions. Additionally, as shown in Figure~\ref{fig:sim_metrics}, examining other similarity metrics shows that the Diff-3DCap model excels, indicating that our approach balances semantic coherence and word completeness in the generated captions.
Although minor weaknesses exist in words overlapping of $\text{BLEU}$, the model can achieve a high similarity score to offset those shortcomings.

% 对Cap3D benchmark的结果分析
Moreover, quantitative results over the Cap3D benchmark are illustrated in Figure~\ref{tb:cap3d_res}.
While our method exhibits lower performance than the large language model-based or GPT-based methods on two semantic metrics, it demonstrated an advantage in the lexical overlap metric.
This can be attributed to the extensive training resources of the large language models, which enable them to attain a deeper understanding of complex geometric features and generate more diverse textual outputs. In contrast, our method emphasizes descriptive vocabulary more closely aligned with similar models.
Moreover, our approach outperform the OpenShape baseline on all metrics, and it can be observed from Figure~\ref{fig:cap3d_res} that the generated captions adequately describe the semantic content of the 3D objects while adopting a coherent narrative style.

\subsection{Ablation study}
\textbf{Parameters.}
Here, we review some significant hyper-parameters of our work.
% render view V
Initially, we examine the impact of varying quantities of 2D projection images on the experimental results. We establish a set of values, \(V \in \{2, 4, 6, 8, 10\}\), and synthesize the generation captions from these diverse viewpoints to generate a comprehensive annotation.
Table~\ref{tb:V} presents a noticeable improvement in the model's performance coupling with increased views. Furthermore, Figure~\ref{fig:ablation_view} illustrates examples of 3D shapes accompanied by captions generated under varying view numbers. Notably, the model fails to produce a readable caption at a view count of 2, repetitively mentioning a single word. This defect underscores the inability of the model to capture essential attributes of shapes, such as material, color, and texture, at lower view counts.
As the view count increases from 4 to 10, a gradual enhancement in the representation of these attributes is observed. This observation proves that an increased number of views enables a more comprehensive representation of 3D objects, which enables the model to generate more accurate descriptions of 3D shapes.
Nevertheless, the limits of our computational resources constrain the exploration of more view settings. Consequently, we employ $V=10$ during subsequent analyses.

% dimension of embedding

\begin{figure}[t]
	\centering
	\includegraphics[width=1\columnwidth]{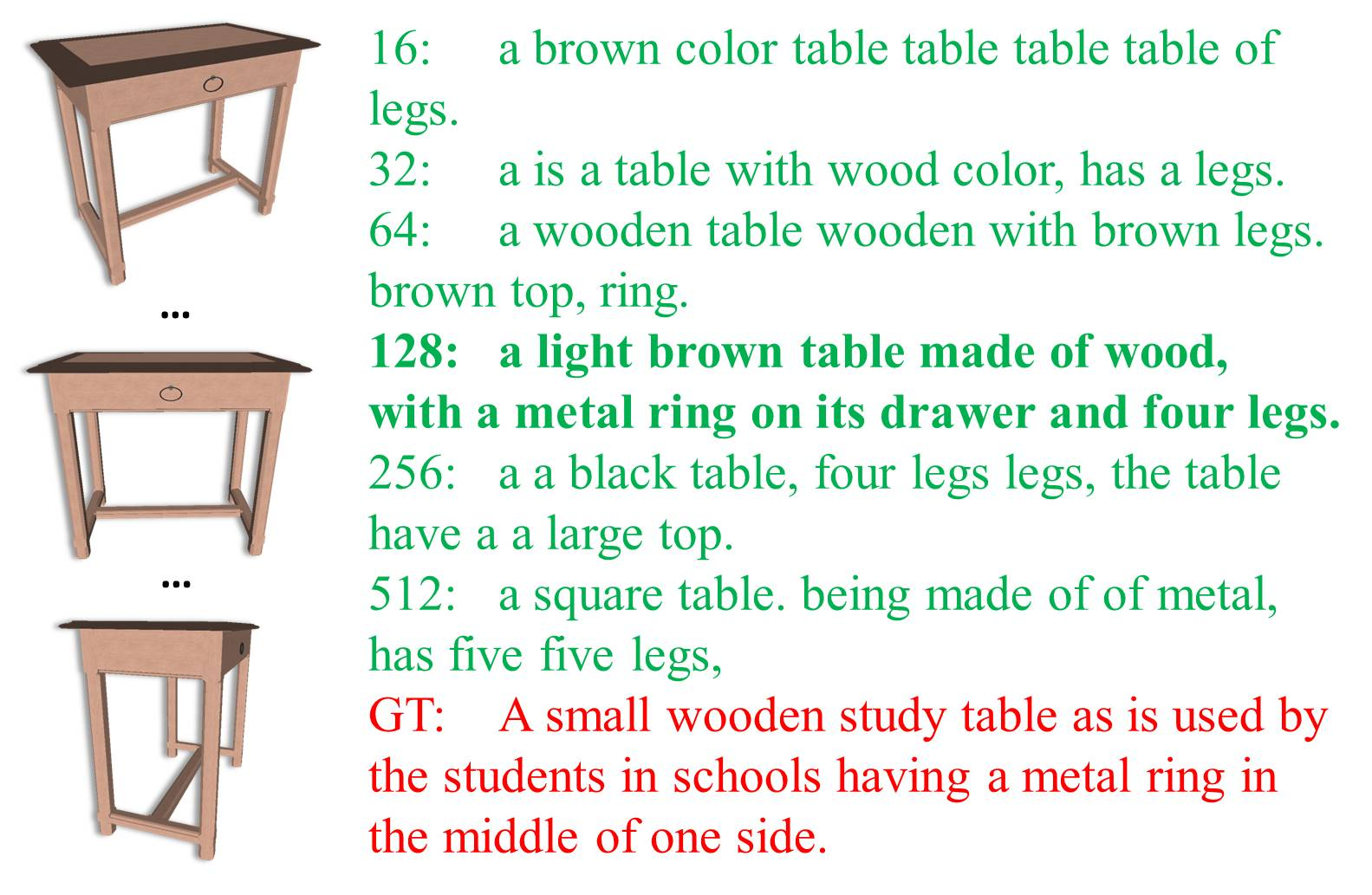}
	\caption{Comparison between generated annotations under different dimensions of embeddings.}
	\label{fig:ablation_dimension}
\end{figure}

Then, we explore the effect of the dimension $H$ of cross-modal VLP embeddings by comparing $H \in \{16, 32, 64, 128, 256, 512\}$. 
Increasing the dimension of the embeddings leads to a more intricate joint representation space, enhancing the richness of information. However, this complexity also results in prolonged training and inference time. As demonstrated in Table~\ref{tb:H} and Figure~\ref{fig:ablation_dimension}, the performance of our Diff-3DCap model shows a consistent improvement across nearly all evaluated metrics as the dimension $H$ is expanded from 16 to 128. When $H$ exceeds 128, there is a decline in metrics performance, which suggests that the model could be susceptible to overfitting at higher dimensions, potentially diminishing its effectiveness on unseen data.

\begin{table*}
	\renewcommand\arraystretch{1.2}
	\begin{threeparttable}
		\caption{The comparison of shape captioning results of our model using different dimensions $H$ of embeddings. Here, $V=10,|S|=5$.}
            \label{tb:H}
		\begin{tabularx}{\textwidth}{*{10}{>{\centering\arraybackslash}X}}
			\toprule
			$H$  &METEOR $\uparrow$ 			  &ROUGE $\uparrow$  				&CIDEr $\uparrow$   			  &BLEU@1 $\uparrow$   	   &BLEU@2 $\uparrow$   	&BLEU@3 $\uparrow$   	 &BLEU@4 $\uparrow$     		     \\
			\midrule
			16    & 0.113          & 0.157           & 0.001 		  & 0.162      & 0.068 & 0.039 & 0.019  \\
			32     & 0.274         & 0.553          & 0.925 		  & 0.716       & 0.703  & 0.715 & 0.694   \\
   			64     & 0.463          &0.597            & 1.004		  & 0.778       & 0.782 & 0.751 & 0.705  \\
			128   &\textbf{0.502}  & \textbf{0.760} 	& \textbf{1.507} & \textbf{0.825}  &\textbf{0.843} &\textbf{0.779} &\textbf{0.758}    \\
			256        & 0.372          & 0.516      & 0.617 		  & 0.740     & 0.693 & 0.484 & 0.413  \\	
			512       & 0.256 		  & 0.501          & 0.523 		  & 0.727       & 0.549 & 0.532 & 0.408   \\
			\bottomrule
		\end{tabularx}
	\end{threeparttable}
\end{table*}

% num of candidate captions S
\begin{table*}
	\renewcommand\arraystretch{1.2}
	\begin{threeparttable}
		\caption{The comparison of shape captioning results of our model using different numbers of candidate samples $|S|$. Here, $V=10, H=128$.}
		\label{tb:S}		
		\begin{tabularx}{\textwidth}{*{10}{>{\centering\arraybackslash}X}}
			\toprule
			$S$  &METEOR $\uparrow$ 			  &ROUGE $\uparrow$  				&CIDEr $\uparrow$   			  &BLEU@1 $\uparrow$   	   &BLEU@2 $\uparrow$   	&BLEU@3 $\uparrow$   	 &BLEU@4 $\uparrow$     		     \\
			\midrule
			1    & 0.198          & 0.279           & 0.042 		  & 0.164      & 0.186 & 0.235 & 0.118  \\
			3        & 0.326          & 0.503      & 0.631 		  & 0.597     & 0.714 & 0.551 & 0.538  \\
			{5}     &\textbf{0.502}  & \textbf{0.760} 	& \textbf{1.507} & \textbf{0.825}  &\textbf{0.843} &\textbf{0.779} &\textbf{0.758}    \\
			\bottomrule
		\end{tabularx}
	\end{threeparttable}
\end{table*}
\begin{figure}[t]
	\centering
	\includegraphics[width=1\columnwidth]{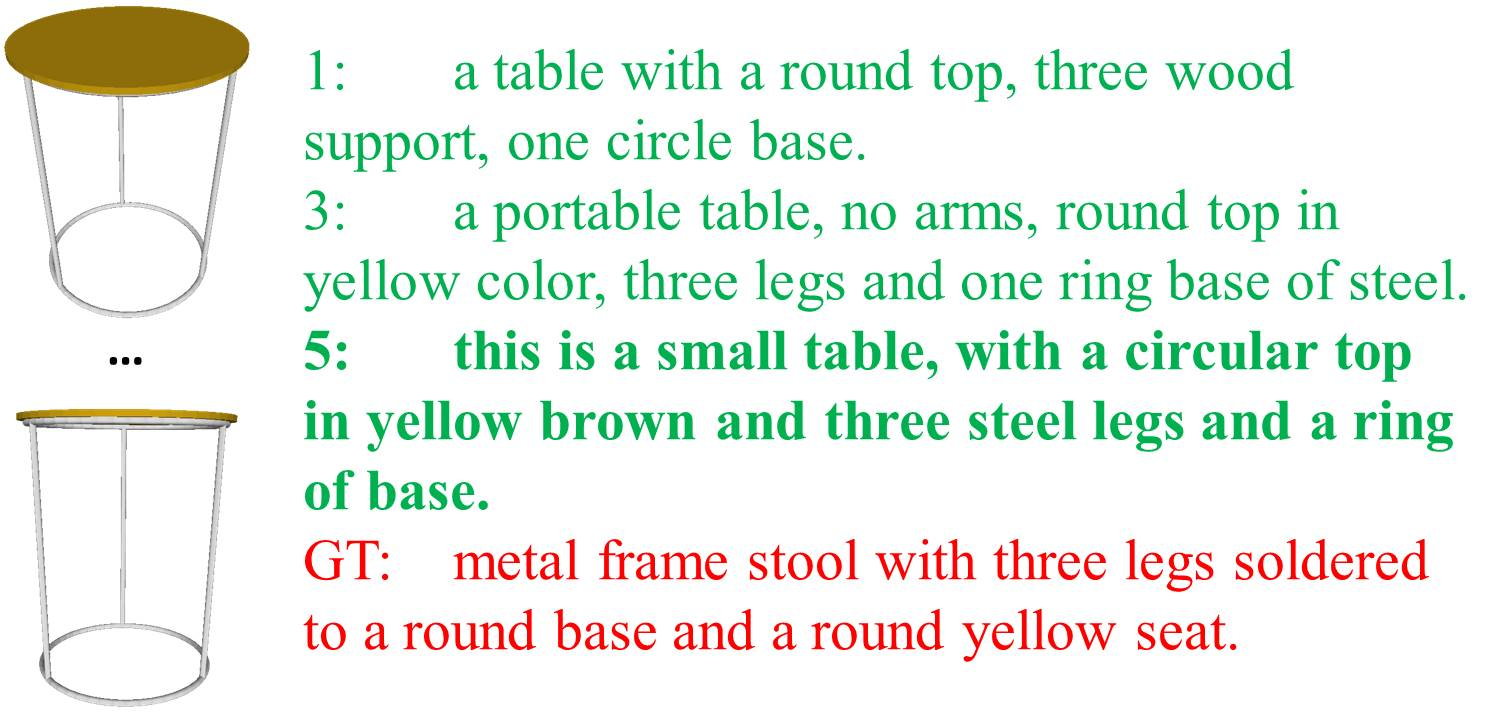}
	\caption{The comparison between generated annotations under different sizes of candidate samples $|S|$.}
	\label{fig:ablation_s}
\end{figure}
Finally, we discuss the variation in the sizes of candidate samples across different projection views.
By utilizing the MBR strategy to consolidate generated candidate captions, we ensure that the caption exhibiting the minor loss is favored because inferior annotations that diverge from the rest are given a more significant loss penalty, reducing their likelihood of being chosen as the ultimate output.
We compare $|S|\in\{1,3,5\}$. As presented in Table~\ref{tb:S} and Figure~\ref{fig:ablation_s}, a noticeable ascending trend exists in the performance metrics as $|S|$ increases from 1 to 5. This pattern suggests that expanding the pool of candidate samples is beneficial for enhancing the quality of the generated output sequences. 
A larger sample size facilitates a more accurate approximation of the model's probability distribution. Consequently, this precision enables the accurate calculation of risk associated with each candidate sample. In this way, we can expect to select the caption with superior quality, as the MBR strategy inherently prefers selections that are not only highly probable but exhibit substantial consistency with other outputs.
We settled on $|S| = 5$, which can not only expedite the inference process but also obtain satisfactory results.
\begin{table}
	\centering
	\begin{threeparttable}
		\caption{The comparison of different visual-language embedding methods on the Cap3D benchmark. ``Ours (OpenShape)" indicates we substitute our pre-trained embedding model with the multi-modal representation of OpenShape.}
		\label{tb:vl_embed}
		\begin{tabularx}{\columnwidth}{
				>{\arraybackslash\hsize=1.29\hsize}X % 加宽Method列
				>{\centering\arraybackslash\hsize=1.18\hsize}X
				>{\centering\arraybackslash\hsize=0.68\hsize}X
				>{\centering\arraybackslash\hsize=0.85\hsize}X
			}
			\toprule
			&Sentence-BERT $\uparrow$  &SimCSE $\uparrow$	&ROUGE-L $\uparrow$ \\
			\midrule
			Ours (OpenShape)     & 39.42 & 39.06 & 20.47  \\	
			Ours          & \textbf{40.10} & \textbf{39.23} & \textbf{21.99}   \\
			\bottomrule
		\end{tabularx}
	\end{threeparttable}
\end{table}
 \begin{figure*}[t]
	\centering
	\includegraphics[width=2\columnwidth]{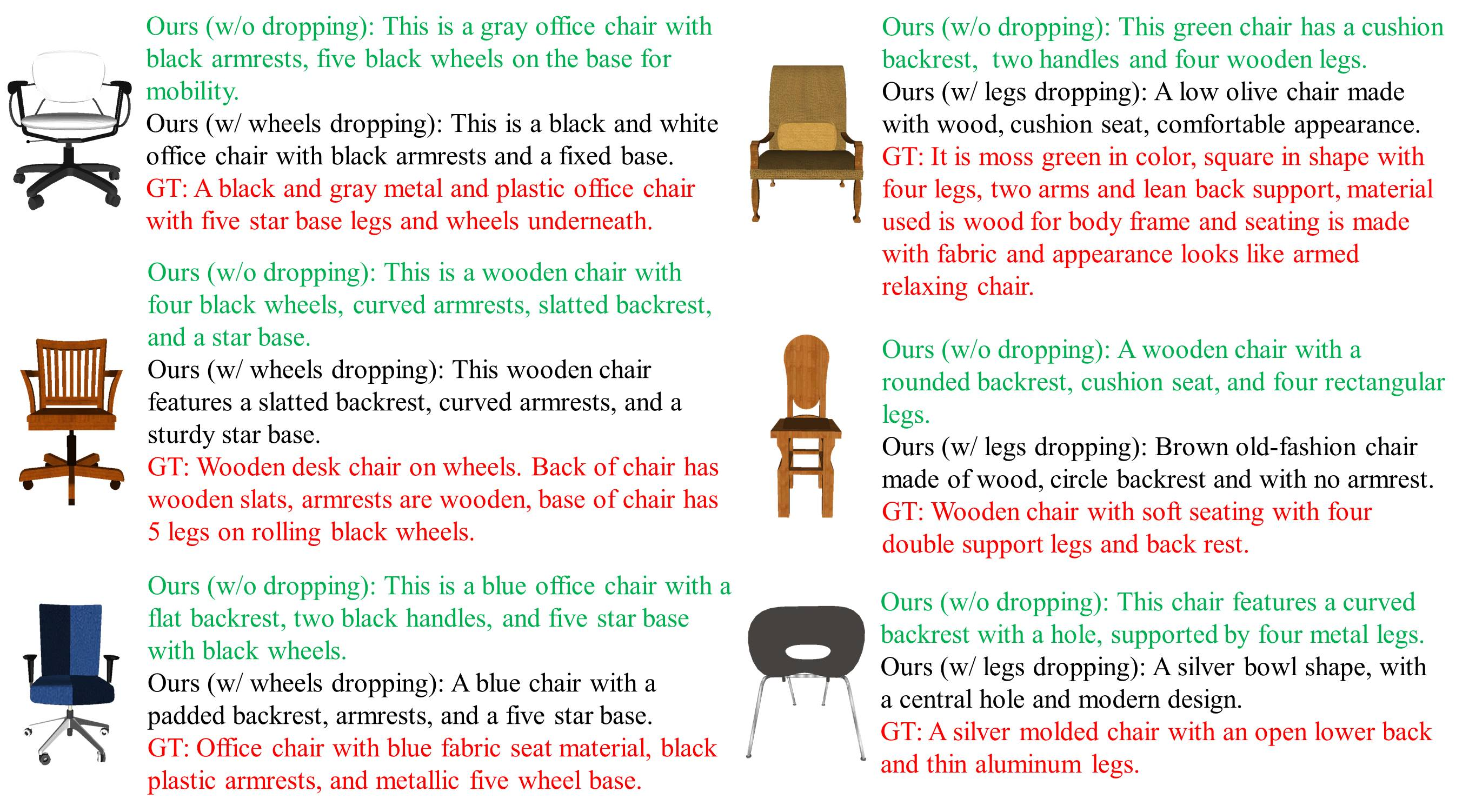}
	\caption{The comparison among results generated by our Diff-3DCap without patch-dropping method(``w/o dropping"), results generated by Diff-3DCap with dropping specific patches for chair objects, and the ground truth captions. The left column illustrates the effect of dropping patches related to wheels(``w/ wheels dropping"), while the right column focuses on the legs(``w/ legs dropping").}
	\label{fig:patch_drop}
\end{figure*}
\begin{figure}[t]
	\centering
	\includegraphics[width=1\columnwidth]{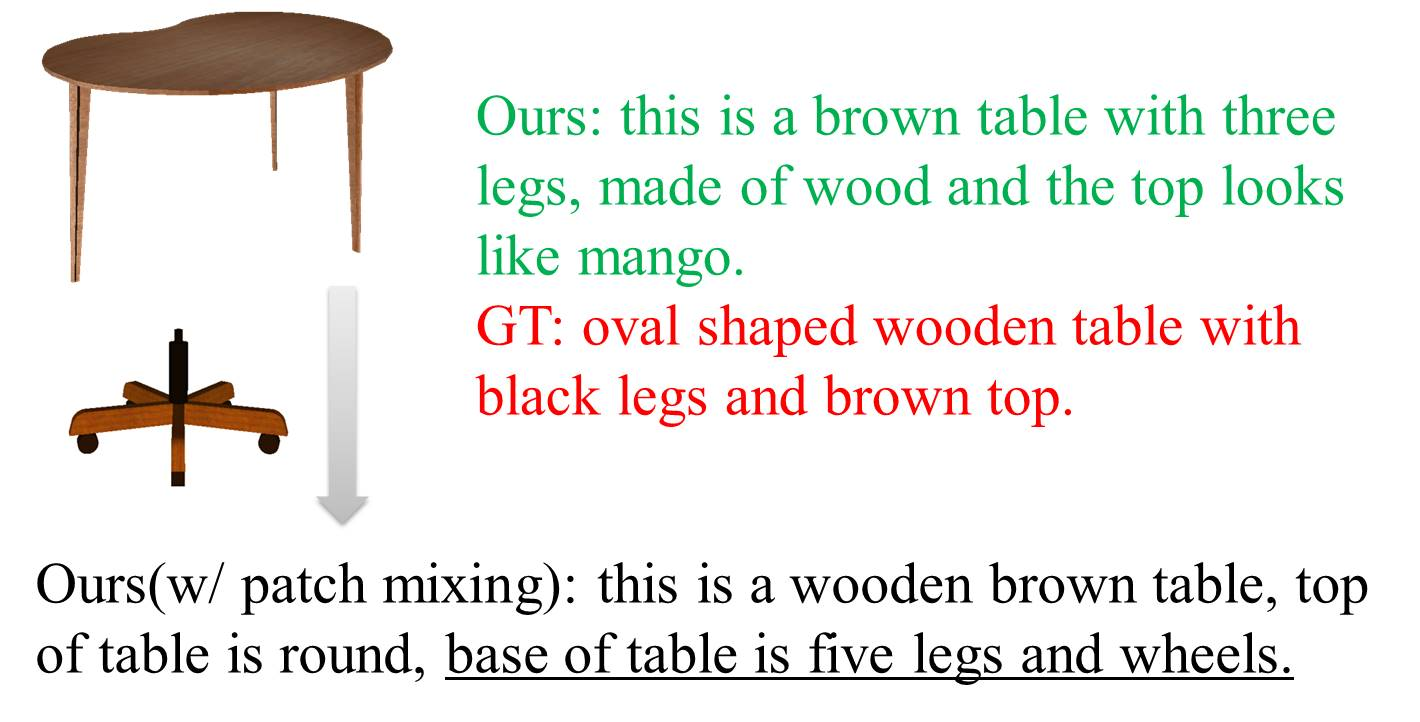}
	\caption{The comparison among results generated by our origin Diff-3DCap, the ground truth captions, and results generated by Diff-3DCap under patch-mixing strategy. The term ``w/ patch mixing" indicates we substitute leg patches of table objects with those of chairs.}
	\label{fig:patch_mix}
\end{figure}
\begin{figure}[t]
	\centering
	\includegraphics[width=1\columnwidth]{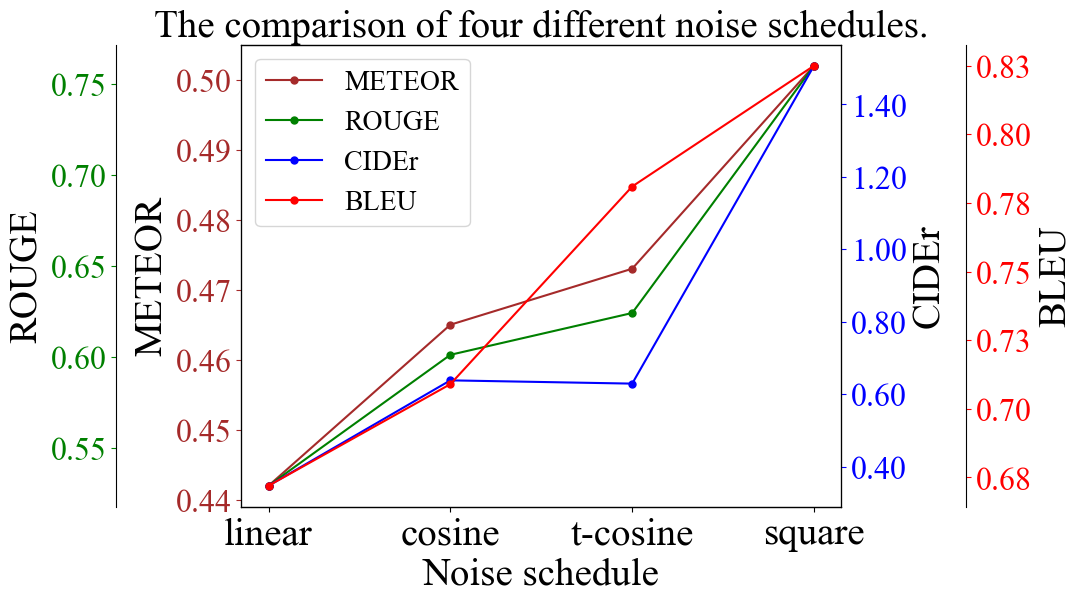}
	\caption{The comparison of noise schedules of the diffusion process, in which t-cosine means truncation cosine noise schedule.}
	\label{fig:noise}
\end{figure}
\begin{figure}[t]
	\centering
	\includegraphics[width=1.0\columnwidth]{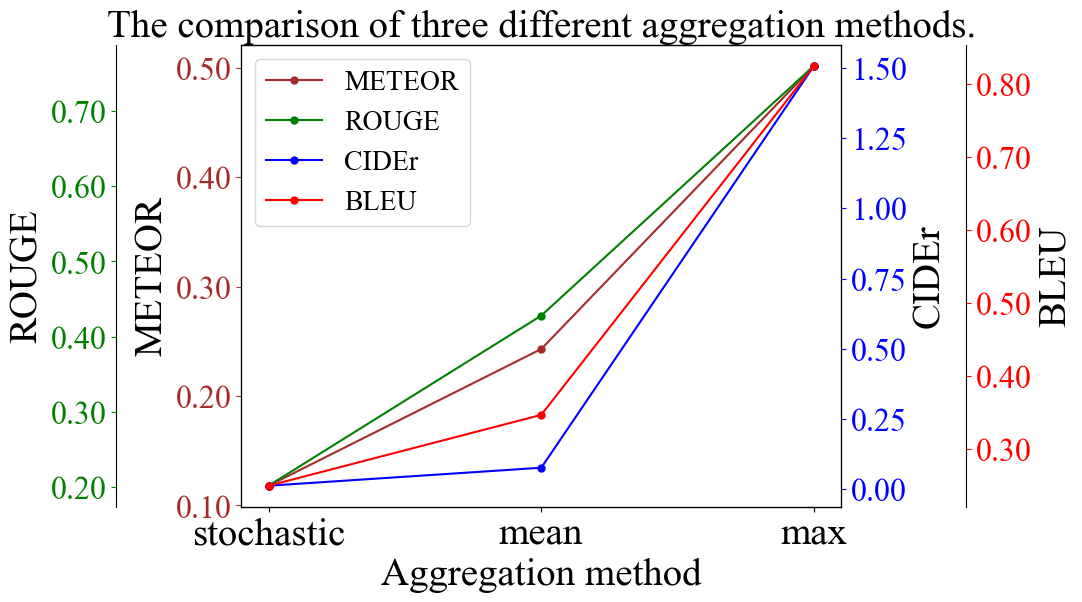}
	\caption{The comparison of different aggregation methods across all projection views.}
	\label{fig:agg_method}
\end{figure}

\textbf{Modules.}
% visual-language embedding method
We initially assess the influence of different visual-language embedding methods on our model performance, comparing our pre-trained embedding model with the OpenShape latent space.
OpenShape trained a 3D point cloud encoder to obtain 3D shape representations that are aligned with a pre-trained CLIP embedding spaces.
As illustrated in Table~\ref{tb:vl_embed}, the implementation of our proposed strategy significantly enhances the quality of captioning results. In contrast to holistic embedding methods, which compress a 3D shape into a single latent vector, our patch-based approach effectively grounds descriptive words to specific structural features. This facilitates a more detailed capture of fine-grained geometric attributes, offering a nuanced understanding of the object's characteristics.

% validate the effectiveness of patch embedding method
We further assess the efficacy of the patch embedding method employed within our visual-language model, particularly focusing on validation through patch mixing and patch dropping techniques.
As illustrated in Figure~\ref{fig:patch_mix}, we implement a patch mixing strategy whereby we interchange leg patches of table objects with those of chair objects before feeding the mixed patches into our captioning pipeline. The results, highlighted in the underlined sentence, demonstrate that the captions generated by our Diff-3DCap model under this patch mixing approach exhibit semantic fusion, effectively merging attributes from a distinct category. 
Additionally, as depicted in Figure~\ref{fig:patch_drop}, we drop specific patches associated with chair objects, including wheels and legs. This selective deletion results in a corresponding lack of descriptive vocabulary for the removed components within the output sentences. 
These results verify that our patch embeddings are capable of encoding part-specific geometric attributes, thereby enhancing the joint learning of geometric features and their corresponding textual descriptions.

% noise schedules
Subsequently, we focus on how different noise schedules affect the model's performance. 
In diffusion models, a noise schedule is crucial for controlling how data is progressively contaminated by noise in the forward diffusion process.
We have evaluated several classical noise scheduling strategies relevant to diffusion models. 
As demonstrated in Figure~\ref{fig:noise}, the outcomes suggest that the square-root noise schedule can yield better results. We assert that the conventional noise scheduling approaches do not adequately accommodate the inherently discrete nature of captions. The square-root noise schedule introduces a higher noise level proximal to $t=0$, facilitating a rapid deviation from the initial state. 
This characteristic enables the model to acquire denoising knowledge more efficiently during the reverse phase, and the accelerated noise injection supports reaching the Gaussian distribution in fewer diffusion steps.

% consolidation method
% stochastic pool,mean pool,max pool
Furthermore, we highlight our view-specific aggregation method for comprehensively describing the 3D objects.
Ideally, the predicted latent variable, denoted as $\hat{\textbf{\emph{x}}}_0^{cap}$, should align with a specific caption within the discrete textual space. We synthesize the embeddings $\hat{\emph{\textbf{x}}}_{0, i}^{cap}$, which is generated from multiple viewpoints with  $i \in \{1, 2,...,V\}$, afterward utilizing a rounding technique following~\cite{li2022diffusion} to identify the word most likely present at each position. Consequently, an effective strategy for view-based aggregation is instrumental in generating high-quality results.
We plan to evaluate the performance of our dataset's three pooling methods: stochastic pooling, mean pooling, and max pooling.
As shown in Figure~\ref{fig:agg_method}, we can observe that max pooling excels.
The effectiveness of max pooling derives from its capability to capture the most significant or crucial features, as specific words or phrases are pivotal for the overall semantic interpretation. Max pooling highlights the most salient elements of the captions, adeptly mimicking the human tendency to focus on critical details in object recognition.
Nevertheless, mean pooling, by averaging all values within a designated scope, risks diminishing these essential features, potentially resulting in embedding vectors that inadequately represent the semantic content of the texts. Moreover, the stochastic pooling method randomly selects a value, which cannot learn enough about prominent expressions and obtain the worst result.

\section{Limitation and future work}\label{lf}
Our approach has several constraints that pave the way for future research and enhancements.
% 三维投影到二维的限制性
Firstly, converting 3D shapes into 2D projection images may simplify or distort the original textures and features, decreasing visual complexity. This simplification subsequently yields textual descriptions that are less detailed and nuanced. Furthermore, using a confined set of perspectives for projection limits our capacity to comprehensively represent the attributes of 3D shapes, thus potentially undermining the depth and accuracy of our annotations.

% 非跨类别
Additionally, we acknowledge a limitation in our model's capability to generalize across different categories. Our approach has been developed with a category-specific focus, undergoing separate training processes for distinct classes, such as tables and chairs. As a result, the model cannot generate text annotations for an unseen category by leveraging the training from another disparate category.

% 预期的后续工作
% 多视图融合
To surmount these limitations and bolster our model's performance in future works, we plan to integrate sophisticated methodologies for multi-view fusion, such as leveraging large language models, to achieve a more comprehensive depiction of 3D shapes. 
% 多类别能力
Furthermore, we are committed to establishing a cross-category training rule to familiarize the model with a broader range of shape categories during its training phase. This initiative is anticipated to enhance the model's capability to generate accurate text annotations for previously unseen categories, thereby significantly extending its applicability and effectiveness.
\section{Conclusion}\label{cl}
% 解决的问题
3D shape captioning is essential in computer graphics and has attracted increasing attention recently.
To further propel the advancement of this field, we propose Diff-3DCap to generate high-quality and varied captions while reducing model complexity. This model employs a view-based approach, leveraging a pre-trained visual-language model to efficiently capture local features by analyzing image patches.
% 主要的特点，列举一二
% 天然的图片指导信号、多视图有效的合并方法
Furthermore, our continuous diffusion model effectively assimilates and processes the latent variables of render images and captions, which is achieved through reliance on a pre-trained visual language model to seamlessly integrate discrete captions into our continuous diffusion framework, negating the need for an additional classifier to provide a guidance signal for text generation. Moreover, our effective consolidation method generates an informative caption with accuracy and semantic similarity over multiple projection views.
% 实验结果
Our experimental result shows that Diff-3DCap can achieve comparable performance against state-of-the-art techniques.

\section*{Acknowledgments}\label{sec:acknowledgements}
This work is supported by the National Natural Science Foundation of China (62172356, 61872321), Zhejiang Provincial Natural Science Foundation of China (LZ25F020012),
%National Key Research and Development Program of China (2017YFB1002600), 
the Ningbo Major Special Projects of the ``Science and Technology Innovation 2025'' (2020Z005, 2020Z007, 2021Z012).

\ifCLASSOPTIONcaptionsoff
  \newpage
\fi

\bibliographystyle{IEEEtran}
\bibliography{reference}

\begin{thebibliography}{10}
\providecommand{\url}[1]{#1}
\csname url@samestyle\endcsname
\providecommand{\newblock}{\relax}
\providecommand{\bibinfo}[2]{#2}
\providecommand{\BIBentrySTDinterwordspacing}{\spaceskip=0pt\relax}
\providecommand{\BIBentryALTinterwordstretchfactor}{4}
\providecommand{\BIBentryALTinterwordspacing}{\spaceskip=\fontdimen2\font plus
\BIBentryALTinterwordstretchfactor\fontdimen3\font minus
  \fontdimen4\font\relax}
\providecommand{\BIBforeignlanguage}[2]{{%
\expandafter\ifx\csname l@#1\endcsname\relax
\typeout{** WARNING: IEEEtran.bst: No hyphenation pattern has been}%
\typeout{** loaded for the language `#1'. Using the pattern for}%
\typeout{** the default language instead.}%
\else
\language=\csname l@#1\endcsname
\fi
#2}}
\providecommand{\BIBdecl}{\relax}
\BIBdecl

\bibitem{chen2019text2shape}
K.~Chen, C.~B. Choy, M.~Savva, A.~X. Chang, T.~Funkhouser, and S.~Savarese,
  ``{Text2Shape}: Generating shapes from natural language by learning joint
  embeddings,'' in \emph{Computer Vision--ACCV 2018: 14th Asian Conference on
  Computer Vision, Perth, Australia, December 2--6, 2018, Revised Selected
  Papers, Part III 14}, 2019, pp. 100--116.

\bibitem{han2019y2seq2seq}
Z.~Han, M.~Shang, X.~Wang, Y.-S. Liu, and M.~Zwicker, ``{Y2Seq2Seq}:
  Cross-modal representation learning for {3D} shape and text by joint
  reconstruction and prediction of view and word sequences,'' in
  \emph{Proceedings of the AAAI Conference on Artificial Intelligence},
  vol.~33, no.~01, 2019, pp. 126--133.

\bibitem{han2020shapecaptioner}
Z.~Han, C.~Chen, Y.-S. Liu, and M.~Zwicker, ``{ShapeCaptioner}: Generative
  caption network for {3D} shapes by learning a mapping from parts detected in
  multiple views to sentences,'' in \emph{Proceedings of the 28th ACM
  International Conference on Multimedia}, 2020, pp. 1018--1027.

\bibitem{luo2024scalable}
T.~Luo, C.~Rockwell, H.~Lee, and J.~Johnson, ``Scalable {3D} captioning with
  pretrained models,'' \emph{Advances in Neural Information Processing
  Systems}, vol.~36, 2024.

\bibitem{deitke2023objaverse}
M.~Deitke, D.~Schwenk, J.~Salvador, L.~Weihs, O.~Michel, E.~VanderBilt,
  L.~Schmidt, K.~Ehsani, A.~Kembhavi, and A.~Farhadi, ``Objaverse: A universe
  of annotated {3D} objects,'' in \emph{Proceedings of the IEEE/CVF Conference
  on Computer Vision and Pattern Recognition}, 2023, pp. 13\,142--13\,153.

\bibitem{li2023blip}
J.~Li, D.~Li, S.~Savarese, and S.~Hoi, ``{BLIP}-2: Bootstrapping language-image
  pre-training with frozen image encoders and large language models,'' in
  \emph{International Conference on Machine Learning}, 2023, pp.
  19\,730--19\,742.

\bibitem{radford2021learning}
A.~Radford, J.~W. Kim, C.~Hallacy, A.~Ramesh, G.~Goh, S.~Agarwal, G.~Sastry,
  A.~Askell, P.~Mishkin, J.~Clark \emph{et~al.}, ``Learning transferable visual
  models from natural language supervision,'' in \emph{International Conference
  on Machine Learning}, 2021, pp. 8748--8763.

\bibitem{achiam2023gpt}
J.~Achiam, S.~Adler, S.~Agarwal, L.~Ahmad, I.~Akkaya, F.~L. Aleman, D.~Almeida,
  J.~Altenschmidt, S.~Altman, S.~Anadkat \emph{et~al.}, ``{GPT}-4 technical
  report,'' \emph{arXiv preprint arXiv:2303.08774}, 2023.

\bibitem{liu2024openshape}
M.~Liu, R.~Shi, K.~Kuang, Y.~Zhu, X.~Li, S.~Han, H.~Cai, F.~Porikli, and H.~Su,
  ``{OpenShape}: Scaling up {3D} shape representation towards open-world
  understanding,'' \emph{Advances in Neural Information Processing Systems},
  vol.~36, 2024.

\bibitem{ho2020denoising}
J.~Ho, A.~Jain, and P.~Abbeel, ``Denoising diffusion probabilistic models,''
  \emph{Advances in Neural Information Processing Systems}, vol.~33, pp.
  6840--6851, 2020.

\bibitem{saharia2022photorealistic}
C.~Saharia, W.~Chan, S.~Saxena, L.~Li, J.~Whang, E.~L. Denton, K.~Ghasemipour,
  R.~Gontijo~Lopes, B.~Karagol~Ayan, T.~Salimans \emph{et~al.},
  ``Photorealistic text-to-image diffusion models with deep language
  understanding,'' \emph{Advances in Neural Information Processing Systems},
  vol.~35, pp. 36\,479--36\,494, 2022.

\bibitem{rombach2022high}
R.~Rombach, A.~Blattmann, D.~Lorenz, P.~Esser, and B.~Ommer, ``High-resolution
  image synthesis with latent diffusion models,'' in \emph{Proceedings of the
  IEEE/CVF Conference on Computer Vision and Pattern Recognition}, 2022, pp.
  10\,684--10\,695.

\bibitem{graikos2024learned}
A.~Graikos, S.~Yellapragada, M.-Q. Le, S.~Kapse, P.~Prasanna, J.~Saltz, and
  D.~Samaras, ``Learned representation-guided diffusion models for large-image
  generation,'' in \emph{Proceedings of the IEEE/CVF Conference on Computer
  Vision and Pattern Recognition}, 2024, pp. 8532--8542.

\bibitem{dhariwal2021diffusion}
P.~Dhariwal and A.~Nichol, ``Diffusion models beat {GAN}s on image synthesis,''
  \emph{Advances in Neural Information Processing Systems}, vol.~34, pp.
  8780--8794, 2021.

\bibitem{li2022diffusion}
X.~Li, J.~Thickstun, I.~Gulrajani, P.~S. Liang, and T.~B. Hashimoto,
  ``Diffusion-{LM} improves controllable text generation,'' \emph{Advances in
  Neural Information Processing Systems}, vol.~35, pp. 4328--4343, 2022.

\bibitem{wu2023ar}
T.~Wu, Z.~Fan, X.~Liu, H.-T. Zheng, Y.~Gong, J.~Jiao, J.~Li, J.~Guo, N.~Duan,
  W.~Chen \emph{et~al.}, ``{AR}-diffusion: Auto-regressive diffusion model for
  text generation,'' \emph{Advances in Neural Information Processing Systems},
  vol.~36, pp. 39\,957--39\,974, 2023.

\bibitem{gong2022diffuseq}
S.~Gong, M.~Li, J.~Feng, Z.~Wu, and L.~Kong, ``{DiffuSeq}: Sequence to sequence
  text generation with diffusion models,'' in \emph{International Conference on
  Learning Representations, ICLR}, 2023.

\bibitem{lin2023text}
Z.~Lin, Y.~Gong, Y.~Shen, T.~Wu, Z.~Fan, C.~Lin, N.~Duan, and W.~Chen, ``Text
  generation with diffusion language models: A pre-training approach with
  continuous paragraph denoise,'' in \emph{International Conference on Machine
  Learning}, 2023, pp. 21\,051--21\,064.

\bibitem{lin2022genie}
Z.~Lin, Y.~Gong, Y.~Shen, T.~Wu, Z.~Fan, C.~Lin, W.~Chen, and N.~Duan,
  ``{GENIE}: Large scale pre-training for text generation with diffusion
  model,'' \emph{arXiv preprint arXiv:2212.11685}, 2022.

\bibitem{ho2022video}
J.~Ho, T.~Salimans, A.~Gritsenko, W.~Chan, M.~Norouzi, and D.~J. Fleet, ``Video
  diffusion models,'' \emph{Advances in Neural Information Processing Systems},
  vol.~35, pp. 8633--8646, 2022.

\bibitem{ho2022imagen}
J.~Ho, W.~Chan, C.~Saharia, J.~Whang, R.~Gao, A.~Gritsenko, D.~P. Kingma,
  B.~Poole, M.~Norouzi, D.~J. Fleet \emph{et~al.}, ``Imagen {Video}: High
  definition video generation with diffusion models,'' \emph{arXiv preprint
  arXiv:2210.02303}, 2022.

\bibitem{kim2023diffusion}
G.~Kim, H.~Shim, H.~Kim, Y.~Choi, J.~Kim, and E.~Yang, ``Diffusion {Video}
  {Autoencoders}: Toward temporally consistent face video editing via
  disentangled video encoding,'' in \emph{Proceedings of the IEEE/CVF
  Conference on Computer Vision and Pattern Recognition}, 2023, pp. 6091--6100.

\bibitem{wu2023tune}
J.~Z. Wu, Y.~Ge, X.~Wang, S.~W. Lei, Y.~Gu, Y.~Shi, W.~Hsu, Y.~Shan, X.~Qie,
  and M.~Z. Shou, ``{Tune-A-Video}: One-shot tuning of image diffusion models
  for text-to-video generation,'' in \emph{Proceedings of the IEEE/CVF
  International Conference on Computer Vision}, 2023, pp. 7623--7633.

\bibitem{10412677}
Y.~Cao, X.~Meng, P.~Y. Mok, T.-Y. Lee, X.~Liu, and P.~Li, ``{AnimeDiffusion}:
  Anime diffusion colorization,'' \emph{IEEE Transactions on Visualization and
  Computer Graphics}, vol.~30, no.~10, pp. 6956--6969, 2024.

\bibitem{10436391}
J.~Xing, M.~Xia, Y.~Liu, Y.~Zhang, Y.~Zhang, Y.~He, H.~Liu, H.~Chen, X.~Cun,
  X.~Wang, Y.~Shan, and T.-T. Wong, ``{Make-Your-Video}: Customized video
  generation using textual and structural guidance,'' \emph{IEEE Transactions
  on Visualization and Computer Graphics}, pp. 1--15, 2024.

\bibitem{anderson2018bottom}
P.~Anderson, X.~He, C.~Buehler, D.~Teney, M.~Johnson, S.~Gould, and L.~Zhang,
  ``Bottom-up and top-down attention for image captioning and visual question
  answering,'' in \emph{Proceedings of the IEEE Conference on Computer Vision
  and Pattern Recognition}, 2018, pp. 6077--6086.

\bibitem{karpathy2015deep}
A.~Karpathy and L.~Fei-Fei, ``Deep visual-semantic alignments for generating
  image descriptions,'' in \emph{Proceedings of the IEEE Conference on Computer
  Vision and Pattern Recognition}, 2015, pp. 3128--3137.

\bibitem{rennie2017self}
S.~J. Rennie, E.~Marcheret, Y.~Mroueh, J.~Ross, and V.~Goel, ``Self-critical
  sequence training for image captioning,'' in \emph{Proceedings of the IEEE
  Conference on Computer Vision and Pattern Recognition}, 2017, pp. 7008--7024.

\bibitem{vinyals2015show}
O.~Vinyals, A.~Toshev, S.~Bengio, and D.~Erhan, ``Show and tell: A neural image
  caption generator,'' in \emph{Proceedings of the IEEE Conference on Computer
  Vision and Pattern Recognition}, 2015, pp. 3156--3164.

\bibitem{rumelhart1986learning}
D.~E. Rumelhart, G.~E. Hinton, and R.~J. Williams, ``Learning representations
  by back-propagating errors,'' \emph{nature}, vol. 323, no. 6088, pp.
  533--536, 1986.

\bibitem{girshick2015fast}
R.~Girshick, ``Fast {R-CNN},'' in \emph{Proceedings of the IEEE International
  Conference on Computer Vision}, 2015, pp. 1440--1448.

\bibitem{ren2016faster}
S.~Ren, K.~He, R.~Girshick, and J.~Sun, ``Faster {R-CNN}: Towards real-time
  object detection with region proposal networks,'' \emph{IEEE Transactions on
  Pattern Analysis and Machine Intelligence}, vol.~39, no.~6, pp. 1137--1149,
  2016.

\bibitem{lu2018neural}
J.~Lu, J.~Yang, D.~Batra, and D.~Parikh, ``Neural baby talk,'' in
  \emph{Proceedings of the IEEE Conference on Computer Vision and Pattern
  Recognition}, 2018, pp. 7219--7228.

\bibitem{huang2019attention}
L.~Huang, W.~Wang, J.~Chen, and X.-Y. Wei, ``Attention on attention for image
  captioning,'' in \emph{Proceedings of the IEEE/CVF International Conference
  on Computer Vision}, 2019, pp. 4634--4643.

\bibitem{lu2017knowing}
J.~Lu, C.~Xiong, D.~Parikh, and R.~Socher, ``Knowing when to look: Adaptive
  attention via a visual sentinel for image captioning,'' in \emph{Proceedings
  of the IEEE Conference on Computer Vision and Pattern Recognition}, 2017, pp.
  375--383.

\bibitem{xu2015show}
K.~Xu, J.~Ba, R.~Kiros, K.~Cho, A.~Courville, R.~Salakhudinov, R.~Zemel, and
  Y.~Bengio, ``Show, attend and tell: Neural image caption generation with
  visual attention,'' in \emph{International Conference on Machine Learning},
  2015, pp. 2048--2057.

\bibitem{yao2018exploring}
T.~Yao, Y.~Pan, Y.~Li, and T.~Mei, ``Exploring visual relationship for image
  captioning,'' in \emph{Proceedings of the European Conference on Computer
  Vision (ECCV)}, 2018, pp. 684--699.

\bibitem{yao2019hierarchy}
{Yao, Ting and Pan, Yingwei and Li, Yehao and Mei, Tao}, ``Hierarchy parsing
  for image captioning,'' in \emph{Proceedings of the IEEE/CVF International
  Conference on Computer Vision}, 2019, pp. 2621--2629.

\bibitem{dosovitskiy2020image}
A.~Dosovitskiy, ``An image is worth 16x16 words: Transformers for image
  recognition at scale,'' \emph{arXiv preprint arXiv:2010.11929}, 2020.

\bibitem{10502235}
Y.~Li, J.~Wang, P.~Aboagye, C.-C.~M. Yeh, Y.~Zheng, L.~Wang, W.~Zhang, and
  K.-L. Ma, ``Visual analytics for efficient image exploration and user-guided
  image captioning,'' \emph{IEEE Transactions on Visualization and Computer
  Graphics}, vol.~30, no.~6, pp. 2875--2887, 2024.

\bibitem{ji2021improving}
J.~Ji, Y.~Luo, X.~Sun, F.~Chen, G.~Luo, Y.~Wu, Y.~Gao, and R.~Ji, ``Improving
  image captioning by leveraging intra-and inter-layer global representation in
  transformer network,'' in \emph{Proceedings of the AAAI Conference on
  Artificial Intelligence}, vol.~35, no.~2, 2021, pp. 1655--1663.

\bibitem{luo2023semantic}
J.~Luo, Y.~Li, Y.~Pan, T.~Yao, J.~Feng, H.~Chao, and T.~Mei,
  ``Semantic-conditional diffusion networks for image captioning,'' in
  \emph{Proceedings of the IEEE/CVF Conference on Computer Vision and Pattern
  Recognition}, 2023, pp. 23\,359--23\,368.

\bibitem{gao2019masked}
J.~Gao, X.~Meng, S.~Wang, X.~Li, S.~Wang, S.~Ma, and W.~Gao, ``Masked
  non-autoregressive image captioning,'' \emph{arXiv preprint
  arXiv:1906.00717}, 2019.

\bibitem{chang2015shapenet}
A.~X. Chang, T.~Funkhouser, L.~Guibas, P.~Hanrahan, Q.~Huang, Z.~Li,
  S.~Savarese, M.~Savva, S.~Song, H.~Su \emph{et~al.}, ``{ShapeNet}: An
  information-rich {3D} model repository,'' \emph{arXiv preprint
  arXiv:1512.03012}, 2015.

\bibitem{luo2024view}
T.~Luo, J.~Johnson, and H.~Lee, ``View selection for {3D} captioning via
  diffusion ranking,'' in \emph{European Conference on Computer Vision}.\hskip
  1em plus 0.5em minus 0.4em\relax Springer, 2024, pp. 180--197.

\bibitem{qi2024shapellm}
Z.~Qi, R.~Dong, S.~Zhang, H.~Geng, C.~Han, Z.~Ge, L.~Yi, and K.~Ma,
  ``{ShapeLLM}: Universal {3D} object understanding for embodied interaction,''
  in \emph{European Conference on Computer Vision}.\hskip 1em plus 0.5em minus
  0.4em\relax Springer, 2024, pp. 214--238.

\bibitem{kim2021vilt}
W.~Kim, B.~Son, and I.~Kim, ``{ViLT}: Vision-and-language transformer without
  convolution or region supervision,'' in \emph{International Conference on
  Machine Learning}, 2021, pp. 5583--5594.

\bibitem{kumar2004minimum}
S.~Kumar and B.~Byrne, ``Minimum {Bayes-Risk} decoding for statistical machine
  translation,'' in \emph{Proceedings of the Human Language Technology
  Conference of the North American Chapter of the Association for Computational
  Linguistics: HLT-NAACL 2004}, 2004, pp. 169--176.

\bibitem{banerjee2005meteor}
S.~Banerjee and A.~Lavie, ``{METEOR}: An automatic metric for mt evaluation
  with improved correlation with human judgments,'' in \emph{Proceedings of the
  {ACL} Workshop on Intrinsic and Extrinsic Evaluation Measures for Machine
  Translation and/or Summarization}, 2005, pp. 65--72.

\bibitem{lin2004rouge}
C.-Y. Lin, ``{ROUGE}: A package for automatic evaluation of summaries,'' in
  \emph{Text Summarization Branches Out}, 2004, pp. 74--81.

\bibitem{vedantam2015cider}
R.~Vedantam, C.~Lawrence~Zitnick, and D.~Parikh, ``{CIDEr}: Consensus-based
  image description evaluation,'' in \emph{Proceedings of the IEEE Conference
  on Computer Vision and Pattern Recognition}, 2015, pp. 4566--4575.

\bibitem{papineni2002bleu}
K.~Papineni, S.~Roukos, T.~Ward, and W.-J. Zhu, ``{BLEU}: A method for
  automatic evaluation of machine translation,'' in \emph{Proceedings of the
  40th Annual Meeting of the Association for Computational Linguistics}, 2002,
  pp. 311--318.

\bibitem{hessel-etal-2021-clipscore}
J.~Hessel, A.~Holtzman, M.~Forbes, R.~Le~Bras, and Y.~Choi, ``{CLIPS}core: A
  reference-free evaluation metric for image captioning,'' in \emph{Proceedings
  of the 2021 Conference on Empirical Methods in Natural Language Processing},
  2021, pp. 7514--7528.

\bibitem{Zhang*2020BERTScore:}
T.~Zhang*, V.~Kishore*, F.~Wu*, K.~Q. Weinberger, and Y.~Artzi, ``{BERTScore}:
  Evaluating text generation with bert,'' in \emph{International Conference on
  Learning Representations}, 2020.

\bibitem{shen2018sequence}
X.~Shen, X.~Tian, J.~Xing, Y.~Rui, and D.~Tao, ``Sequence-to-sequence learning
  via shared latent representation,'' in \emph{Proceedings of the AAAI
  Conference on Artificial Intelligence}, vol.~32, no.~1, 2018.

\bibitem{song2018cross}
Y.~Song and M.~Soleymani, ``Cross-modal retrieval with implicit concept
  association,'' \emph{arXiv preprint arXiv:1804.04318}, 2018.

\bibitem{venugopalan2015sequence}
S.~Venugopalan, M.~Rohrbach, J.~Donahue, R.~Mooney, T.~Darrell, and K.~Saenko,
  ``Sequence to sequence-video to text,'' in \emph{Proceedings of the IEEE
  International Conference on Computer Vision}, 2015, pp. 4534--4542.

\bibitem{hong20233d}
Y.~Hong, H.~Zhen, P.~Chen, S.~Zheng, Y.~Du, Z.~Chen, and C.~Gan, ``{3D-LLM}:
  Injecting the {3D} world into large language models,'' \emph{Advances in
  Neural Information Processing Systems}, vol.~36, pp. 20\,482--20\,494, 2023.

\bibitem{hurst2024gpt}
A.~Hurst, A.~Lerer, A.~P. Goucher, A.~Perelman, A.~Ramesh, A.~Clark, A.~Ostrow,
  A.~Welihinda, A.~Hayes, A.~Radford \emph{et~al.}, ``{GPT}-4o system card,''
  \emph{arXiv preprint arXiv:2410.21276}, 2024.

\end{thebibliography}


% references sectionIEEEtran}
%\bibliography{mybibfile}
%\newpage
%\bibliographystyle{

\begin{IEEEbiography}[{\includegraphics[width=1in,height=1.25in,clip,keepaspectratio]{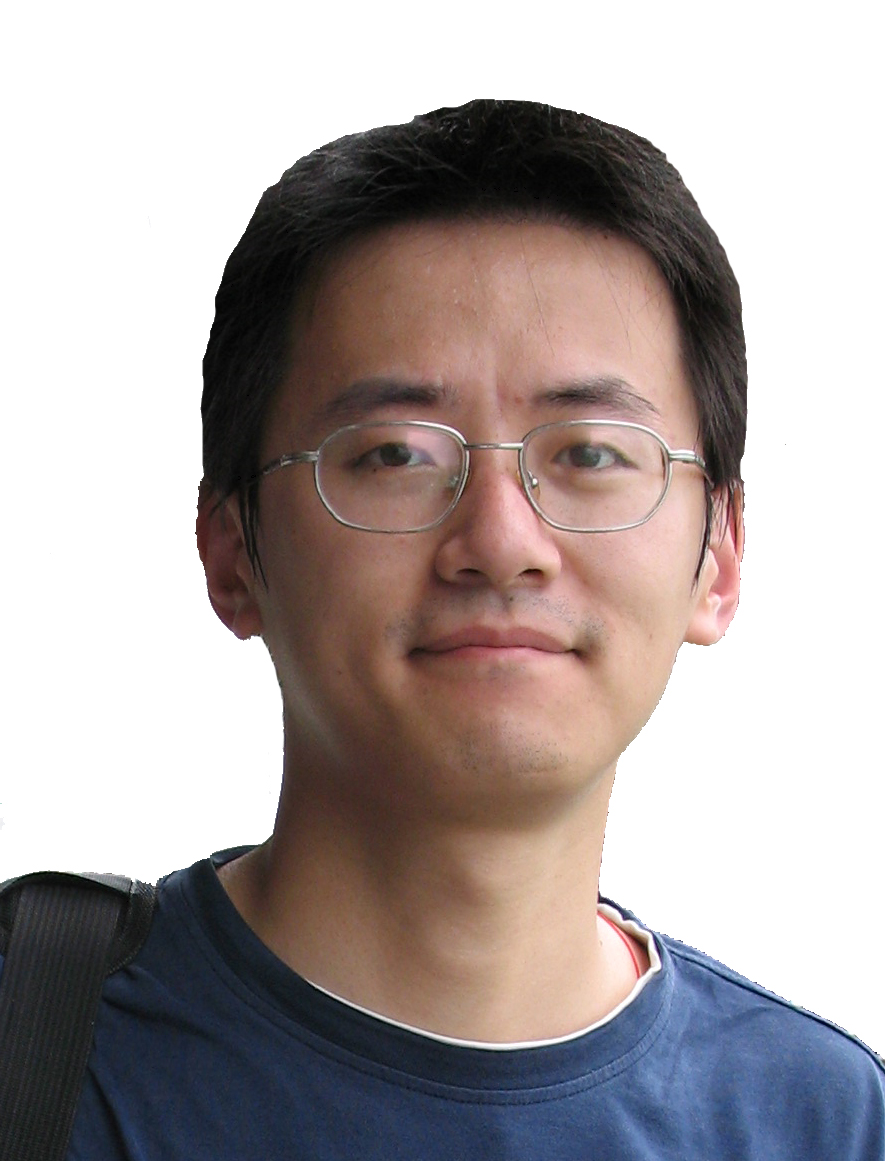}}]{Zhenyu Shu}
got his Ph.D. degree in 2010 at Zhejiang University, China. He is now working as a full professor at NingboTech University. His research interests include computer graphics, digital geometry processing and machine learning. He has published over 30 papers in international conferences or journals.
\end{IEEEbiography}

\begin{IEEEbiography}[{\includegraphics[width=1in,height=1.25in,clip,keepaspectratio]{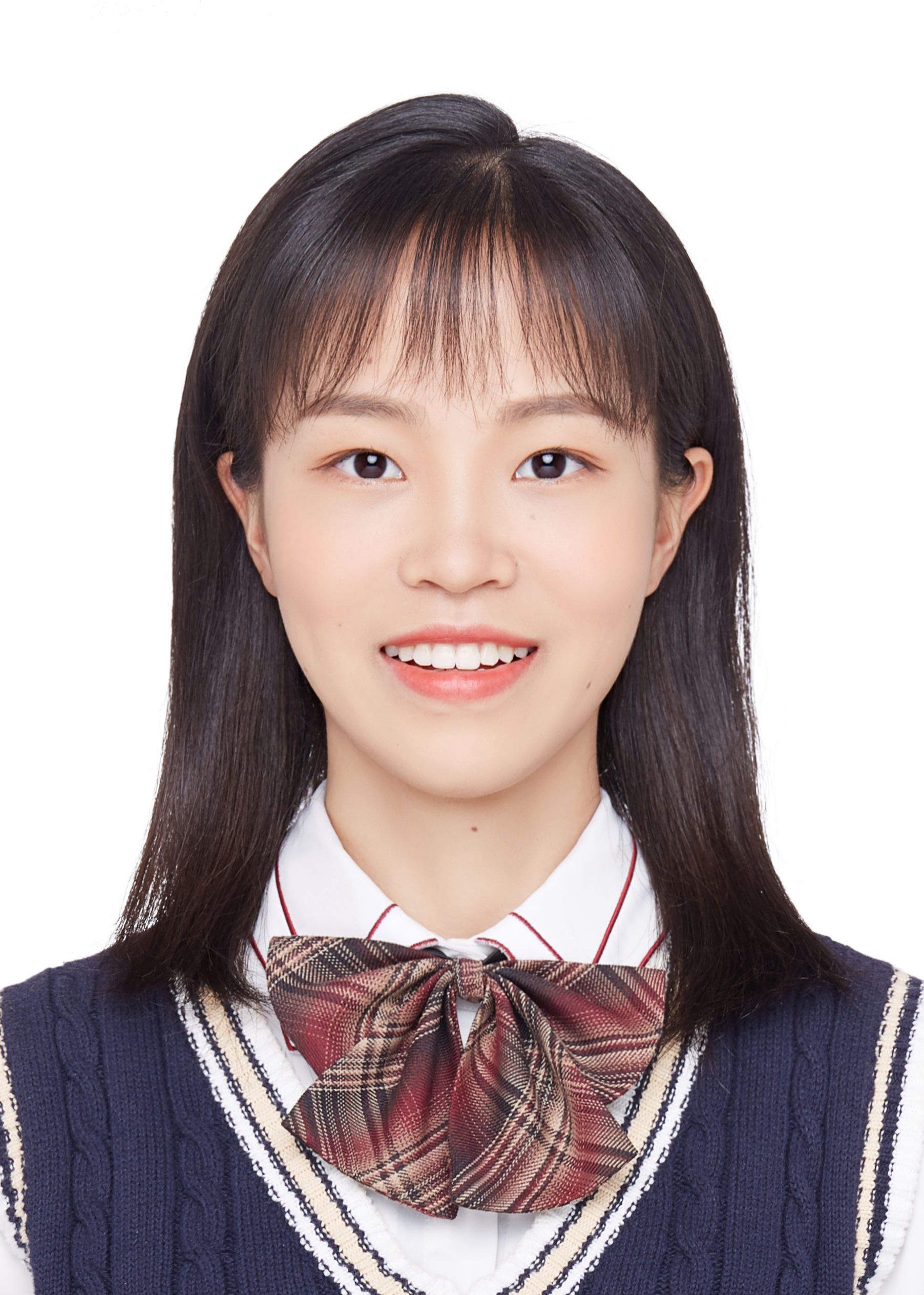}}]{Jiawei Wen}
	 is a graduate student of the College of Computer Science and Technology at Zhejiang University. Her research interests include computer graphics and machine learning.
\end{IEEEbiography}

\begin{IEEEbiography}[{\includegraphics[width=1in,height=1.25in,clip,keepaspectratio]{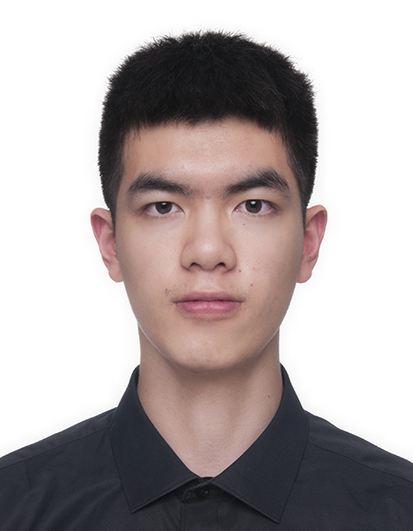}}]{Shiyang Li} is a graduate student of the College of Computer Science and Technology at Zhejiang University. His research interests include computer graphics and machine learning.
\end{IEEEbiography}

\begin{IEEEbiography}[{\includegraphics[width=1in,height=1.25in,clip,keepaspectratio]{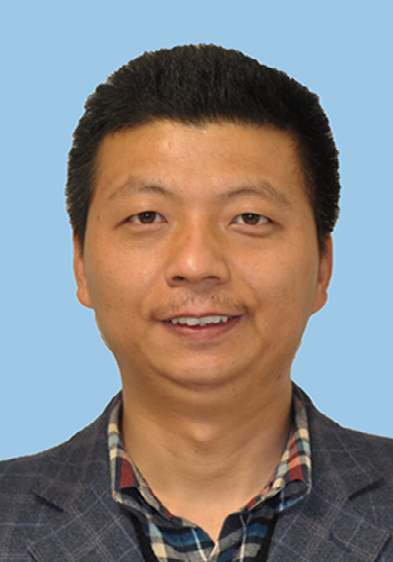}}]{Shiqing Xin}
	is a professor at the Faculty of School of Computer Science and Technology in Shandong University. He received his Ph.D. degree in applied mathematics at Zhejiang University in 2009. His research interests include computer graphics, computational geometry and 3D printing.
\end{IEEEbiography}

\begin{IEEEbiography}[{\includegraphics[width=1in,height=1.25in,clip,keepaspectratio]{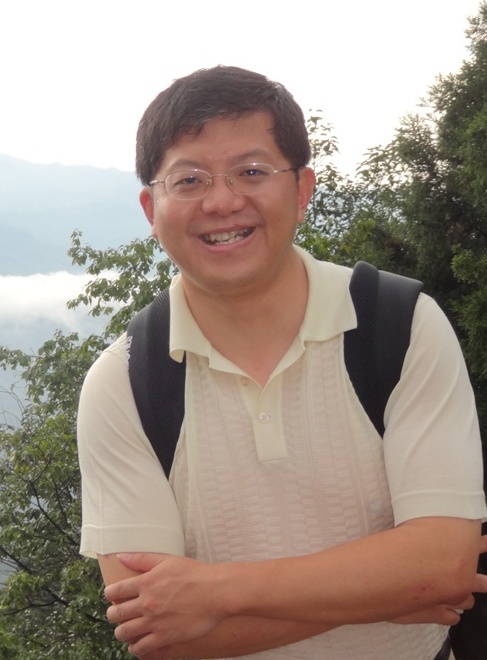}}]{Ligang Liu} received the BSc degree in 1996 and the Ph.D. degree in 2001 from Zhejiang University, China. He is a professor at the University of Science and Technology of China. Between 2001 and 2004, he was at Microsoft Research Asia. Then he was at Zhejiang University during 2004 and 2012. He paid an academic visit to Harvard University during 2009 and 2011. His research interests include geometric processing and image processing. He serves as the associated editors for journals of IEEE Transactions on Visualization and Computer Graphics, IEEE Computer Graphics and Applications, Computer Graphics Forum, Computer Aided Geometric Design, and The Visual Computer. His research works could be found at his research website: http://staff.ustc.edu.cn/lgliu
\end{IEEEbiography}

\vfill


%\newpage



%\vfill


\end{document}